\documentclass[aps,groupedaddres8s,fleqn,nofootinbib,twocolumn]{revtex4}
\usepackage{graphicx}
\usepackage{xcolor}
\usepackage{float}
\usepackage{amsmath,amssymb}
\usepackage{float}
\usepackage{amsfonts}
\usepackage{verbatim}
\usepackage{flushend}
\usepackage{balance}
\usepackage{endnotes}
\usepackage{footnote}
\usepackage{adjustbox}
\usepackage{gensymb}
\usepackage{float}
\usepackage{epigraph}
\usepackage{xcolor}
\usepackage[utf8]{inputenc}
\usepackage{setspace}

\newcommand{\beq}{\begin{eqnarray}}
\newcommand{\eeq}{\end{eqnarray}}
\usepackage{amsmath}

\begin{document}

\title{Generally applicable physics-based equation of state for liquids}
\author{J. E. Proctor$^{\star1}$}
\address{$^1$ Materials and Physics Research Group, University of Salford, Manchester M5 4WT, UK}
\author{K. Trachenko$^{\star2}$}
\address{$^2$ School of Physical and Chemical Sciences, Queen Mary University of London, Mile End Road, London, E1 4NS, UK}

\address{$\star$ j.e.proctor@salford.ac.uk, k.trachenko@qmul.ac.uk}

\begin{abstract}
Physics-based first-principles pressure-volume-temperature equations of state (EOS) exist for solids and gases but not for liquids due to the long-standing fundamental problems involved in liquid theory. Current EOS models that are applicable to liquids and supercritical fluids at liquid-like density under conditions relevant to planetary interiors and industrial processes are complex empirical models with many physically meaningless adjustable parameters. Here, we develop a generally applicable physics-based (GAP) EOS for liquids including supercritical fluids at liquid-like density. The GAP equation is explicit in the internal energy, and hence links the most fundamental macroscopic static property of fluids, the pressure-volume-temperature EOS, to their key microscopic property: the molecular hopping frequency or liquid relaxation time, from which the internal energy can be obtained. We test our GAP equation against available experimental data in several different ways and find good agreement. Our GAP equation, unavoidably and similarly to solid EOS, contains a semi-empirical term giving the energy of the static sample as a function of volume only ($E_{ST}(V)$).  Our testing includes studies along isochores, in order to examine the validity of the GAP equation independently of the validity of any function we may choose to utilize for $E_{ST}(V)$.  The only other adjustable parameter in the equation is the Grüneisen parameter for the fluid. We observe that the GAP equation is similar to the Mie-Gr\"{u}neisen solid EOS in a wide range of the liquid phase diagram. This similarity is ultimately related to the condensed state of these two phases. On the other hand, the differences between the GAP equation and EOS for gases are fundamental. Finally, we identify the key gaps in the experimental data that need to be filled in to proceed further with the liquid EOS.
\end{abstract}

\maketitle

\tableofcontents

\section{Introduction}

\subsection{Liquid theory}

Problems involved in liquid theory were long considered to be of fundamental nature and unworkable. As discussed by Landau, Lifshitz and Pitaevskii (LLP), these problems are (a) strong interatomic interactions combined with dynamical disorder and (b) no small parameter in the theory \cite{landaustat,pitaevskii,ahiezer}. As a result, for a long time no general thermodynamic theory of liquids was thought to be feasible, in contrast to theories of solids and gases developed over a century ago \cite{mybook}. Whereas calculating generally-applicable thermodynamic properties such as energy and heat capacity and their temperature dependence have become an essential part of theories of solids and gases, deriving such general relations was ruled out in liquids \cite{landaustat,pitaevskii} (according to Peierls \cite{peierls-frenkel}, Landau had always maintained that developing a theory of liquids was impossible).

These fundamental problems apply to all theoretical approaches to liquids based on considering liquid interactions and structure explicitly, which has been the predominant approach in statistical physics of liquids in the last century \cite{borngreen,green2,kirkwoodbook,fisher,barkerhenderson,egelstaff,hansen1,hansen2,march,faber,wca1,wca2,rosentar,zwanzig}. These approaches include expanding interactions into short-range repulsive reference and attractive terms in simple models (see, for example, Refs. \cite{wca1,wca2,rosentar,zwanzig}): the expansions and coefficients in these expansions remain system-dependent and the results remain not generally applicable.

The first part of the problem stated by LLP can be illustrated by writing the liquid energy in terms of the concentration $n$, pair distribution function $g(r)$ and interaction potential $u(r)$ (the "energy equation"):

\begin{equation}
E=\frac{3}{2}NT+\frac{n}{2}\int g(r)u(r)dV
\label{enint}
\end{equation}

Analogous equations exist to give the pressure and bulk modulus in terms of $g(r)$ and $u(r)$ (reviewed in \cite{proctor2}). $k_{\rm B}=1$ here, and throughout the theory section.  The term "liquid" refers, here and subsequently, to liquids in the subcritical regime and to supercritical fluids at liquid-like density.

Early theories \cite{kirkwoodbook,borngreen,zwanzig,barkerhenderson} considered the goal of statistical physics of liquids to be to start with liquid structure and intermolecular interactions such as $g(r)$ and $u(r)$ in Eq. \eqref{enint} and on this basis work out liquid thermodynamic properties. Working towards this goal involved developing analytical models for liquid structure and interactions, and this has become the essence of liquid theories \cite{borngreen,green2,kirkwoodbook,fisher,barkerhenderson,egelstaff,hansen1,hansen2,march,faber,wca1,wca2,rosentar,zwanzig}. The problem with this approach is the one stated by LLP: the interaction $u(r)$ in liquids is both strong and system-specific, hence $E$ in Eq. \eqref{enint} is strongly system-dependent. It was for this reason that no generally applicable theory of liquids was considered possible \cite{landaustat,pitaevskii}.

An additional problem is that interactions and correlation functions are generally not available apart from simple models and can be generally complex involving many-body, hydrogen-bond interactions and so on. This precludes calculation of the liquid energy in theories based on Eq. \eqref{enint} or its extensions involving higher-order correlation functions \cite{borngreen,barkerhenderson}. Even when $g(r)$ and $u(r)$ are available in simple cases, the calculation involving Eq. \eqref{enint} or similar is not enough: in order to explain experimental data such as heat capacity of real liquids \cite{ropp,wallacecv,wallacebook,proctor1,proctor2,chen-review}, one still needs to develop a physical model.

The small parameter in solids simplifying theory are small atomic displacements from equilibrium positions, but this ostensibly does not apply to liquids because liquids do not have stable equilibrium points that can be used to sustain these small phonon displacements. Weakness of interactions assumed in the theory of gases can not apply to liquids either because interactions in liquids, the condensed state of matter, are as strong as in solids. This is the second, no small parameter problem, stated by LLP.

Contrary to what was commonly believed (see, e.g., Ref. \cite{rosenscaling}) and differently to what is often currently considered, popular models used to discuss liquids are inapplicable to understanding the most important properties of real liquids such as energy and heat capacity. These models include the widely used Van der Waals model, the hard-spheres model and their extensions \cite{hansen2,ziman,march,parisihard,parisi}. Both models give the specific heat $c_v=\frac{3}{2}k_{\rm B}$ \cite{landaustat,wallacecv,wallacebook} and hence describe the non-interacting ideal gas from thermodynamic point of view. This is far removed from experiments showing liquid $c_v=3k_{\rm B}$ close to melting \cite{wallacecv,wallacebook,proctor2,nist,ropp,mybook} where this $c_v$ is as large as in solids. Properties of gases and solids {\it are} very different due to interactions \cite{landaustat,landaustat1}, hence models that give $c_v=\frac{3}{2}k_{\rm B}$ instead of $c_v=3k_{\rm B}$ miss essential physics. We note that simple models above were also used as reference states to calculate the energy \eqref{enint} by expanding interactions into attractive and repulsive parts (see, e.g., Refs. \cite{barkerhenderson,wca1,wca2,zwanzig,rosentar}). These parts understandably play different roles at low and high density, however this method faces the same problem outlined by LLP: the interactions and expansion coefficients are still strongly system-dependent and so are the final results, precluding a general theory.

As a result of these fundamental problems, understanding liquid thermodynamic properties (both their actual values and temperature dependence) theoretically has remained a long-standing problem. This remained a serious gap in both research and teaching \cite{chen-review,granato,prescod}.

Notably, the above problems do not originate in the solid state theory. The fundamental principle of statistical physics is that properties and essential physics of an interacting system are governed by its excitations \cite{landaustat,landaustat1}. We are able to readily apply this principle to solids, both crystalline and amorphous, where these excitations are collective excitations, phonons. The theory based on phonons is physically transparent, predictive, generally applicable and forms the cornerstone of the solid state theory as well as other areas. Remarkably, there is no need to explicitly consider structure and interactions of solids to understand their basic properties. Most important results such as the universal temperature dependence of energy and heat capacity of solids readily come out in the phonon theory \cite{landaustat,landaustat1}. This also applies to deriving the equations of state for solids \cite{anderson1}.

It is therefore interesting to observe that the approach to the liquid theory diverged from the solid state theory in its fundamental perspective: the liquid theories were based on correlation functions and interatomic interactions, whereas the solid state theory operated in terms of phonons. However, there were notable exceptions. Sommerfeld and Brillouin \cite{brillouin,somm,br1,br2,brilprb} considered that the liquid energy and thermodynamic properties are fundamentally related to phonons as in solids and sought to discuss liquid properties on the basis of a modified Debye theory. These ideas were published starting from 1913 and shortly after the Einstein and Debye theories of solids \cite{einstein,debyepaper} were published laying the foundations of the solid state theory. Trying to link liquid thermodynamics to phonons has extended over a substantial period of Brillouin's research. Apart from isolated attempts \cite{wannier,faber,wallacebook}, this line of enquiry has stopped, and liquid theories based on structure and interactions in Eq. \eqref{enint} were pursued instead. Whereas the Debye and Einstein theories have become part of every textbook on solids, a theory of liquid thermodynamics has remained unworkable for about a century that followed. An important reason is that, differently from solids, the nature of phonons in liquids remained unclear for a long time \cite{mybook}.

\subsection{Equation of state}

The above problems of liquid theory have understandably impacted the development of physics-based first-principles liquid equation of state (EOS). These equations exist for solids and gases but not for liquids.

At present there exists no standard source of fluid $PVT$ EOS data for planetary science or industrial applications of liquids and supercritical fluids, and $PVT$ EOS are based on complex empirical models for industrial applications of simple and complex liquids.  In the remainder of the introduction we will outline the nature of the problems with existing approaches, and our proposed solution.

At this stage, it is appropriate to define the nomenclature relating to EOS and liquids for clarity throughout the remainder of this work.  The term "equation of state" is formally defined as any equation linking the values of two or more state variables.  Often these variables are assumed to be pressure and volume, or pressure, volume and temperature.  In the present work we need to discuss EOS linking various combinations of state variables, so will explicitly list these in all cases.  Using this nomenclature, the ideal gas EOS is a $PVT$ EOS, the Birch-Murnaghan EOS is a $PV$ EOS and the law giving the internal energy of a monatomic ideal gas $E=(3/2)RT$ is an $ET$ EOS.  The only exceptions to this nomenclature will be when we describe the fundamental EOS which can provide outputs for various state variables from a single equation for the Helmholtz free energy and the Mie-Gr{\"u}neisen EOS for solids as the M-G EOS. The term "liquid" will be used to describe the liquid state in the subcritical regime, as well as the part of the supercritical fluid state on the liquid-like (high density) side of the Frenkel line \cite{flreview}.  The term "particle" will be used to refer to the atoms comprising an atomic fluid, and the molecules comprising a molecular fluid.

The first EOS to be proposed that explicitly (albeit qualitatively) accounts for the balance between attraction and repulsion between fluid particles that characterizes a liquid was the van der Waals $PVT$ EOS, referred to as a "cubic" EOS because it can be written as a cubic polynomial in $V$:

\begin{equation}
P=\frac{RT}{V-b}-\frac{a}{V^2}
\label{vderw}
\end{equation}

Since then a wide variety of other cubic $PVT$ EOS (e.g. Patel-Teja, Redlich-Kwong, Peng-Robinson etc.) have been proposed based on the van der Waals $PVT$ EOS but incorporating an additional- $V$-dependent term and some form of temperature dependence for the van der Waals parameter $a$.  These EOS can be written in the following form (referred to as the generalized cubic $PVT$ EOS)\cite{deiterskraska,proctor2}:

\begin{equation}
P=\frac{RT}{V-b}-\frac{a(T)}{V^2+cV+d}
\label{gencub}
\end{equation}

However, these equations are essentially the ideal gas EOS with corrections so do not perform well for accurate modelling of liquid properties.

The current state of the art in liquid and supercritical fluid $PVT$ EOS for use in industry is a methodology called the fundamental (or reference) EOS \cite{proctor2,deiterskraska,setzmann}.  The fundamental EOS for a fluid is an equation giving the Helmholtz free energy of the fluid as a function of molar volume and temperature ($F$($V$,$T$)).  Most macroscopic static and dynamic properties can be obtained from derivatives of this function, such as the speed of sound and the internal energy.  This method ensures that the obtained values of different fluid properties are at least consistent, even if they are not necessarily correct.  As far as $PVT$ EOS are concerned, these can be obtained by calculating the pressure from $F$($V$,$T$) and thus linking the pressure to the $V,T$ input variables.  Thus various different EOS can be obtained from a single fundamental EOS, and will henceforth be referred to as "output" from the fundamental EOS.  Fundamental EOS are available for most simple fluids and many complex fluids, albeit in some cases over a limited $V,T$ range.  Whilst the fundamental EOS for most fluids are described in publications (for instance \cite{setzmann, tegelerArEOS, spanN2EOS}), using the fundamental EOS usually involves generating whatever output EOS is required using a software package.  In this work we use the NIST webbook\cite{nist} and the ThermoC software\cite{thermoc} to generate the required outputs from the fundamental EOS.

Whilst useful, the fundamental equations for $F$($V$,$T$) are almost entirely empirical, and typically incorporate ca. 50 dimensionless and physically meaningless free fitting parameters.  The process of fitting to the available experimental data (linear and non-linear regression analysis) often involves arbitrary choices of which data to weight more heavily in the fitting process, and can lead to a fundamental EOS which overfits to the large amount of data available in the gas state and critical region at the expense of the more sparse data at liquid-like density at high temperature.  Due to the overfitting, the fundamental EOS typically provides outputs that match most of the experimental data very well, interpolates between the data in an effective manner, and allows disparate sources of experimental data on fluids to be collated and checked for mutual consistency.  In our earlier work \cite{proctor1} we used the fundamental EOS viscosity and internal energy output along isochores extensively instead of the original experimental data, as we would otherwise have had to interpolate between the experimental datapoints ourselves to obtain their values along the same isochore.  However, extrapolation of the fundamental EOS output to $VT$ conditions at which it was not fitted to experimental data has been shown to be unreliable in many cases (Ref. \cite{proctor2} and refs. therein).  This is due to the empirical nature of the fundamental EOS for $F$($V$,$T$).  Care is required to distinguish between interpolation and extrapolation when generating output from the fundamental EOS as the $PT$ range of experimental data advertised in the abstract of publications presenting the fundamental EOS for different fluids is usually that of the $PVT$ data.  Dynamic properties such as heat capacity and speed of sound are often fitted over a far smaller $PT$ range and even the $PVT$ data are often very limited at liquid-like densities above 300 K.  Transport properties such as viscosity are fitted using a separate mainly empirical model\cite{lemmonvisc}.

More recent work explored scaling arguments simplifying the description of properties on the phase diagram (see, e.g. Refs. \cite{scaling1,scaling2,scaling3,scaling4,bell1,bell2}). This includes using the Rosenfeld conjecture \cite{rosenscaling} that certain liquid properties scale with the excess entropy. This was based on using the hard-sphere model mentioned earlier. We will return to this conjecture in Section \ref{liquidexc}.

In the last couple of decades a set of new results have emerged related to phonons in liquids, allowing the new approach to liquid $PVT$ EOS outlined in this work. It has taken a combination of new experiments, theory and modelling to understand phonons in liquids well enough to connect them to liquid thermodynamic properties. Recall that it is this connection between thermodynamic properties and phonons which formed the basis of the Einstein and Debye approach to solids and laid the foundations of the modern solid state theory. This, in turn, enabled the development of physics-based EOS for solids discussed in Section \ref{eossol}.

The upshot is that the liquid theory can too be developed on the basis of phonons.  The key point is that, differently from solids, the phase space available to phonons in liquids is not fixed but is instead {\it variable}. In particular, this phase space reduces with temperature.  This reduction quantitatively explains the experimental liquid data and in particular the decrease of liquid specific heat from the solidlike to the ideal gas value with increasing temperature \cite{ropp,mybook,proctor1,proctor2}.

The small parameter in this liquid theory is therefore the same as in the solid state theory: small phonon displacements. However, in important difference to solids, this small parameter operates in a variable phase space. This addresses the problems stated by Landau, Lifshitz and Pitaevskii above. As in the solid theory, the phonon gas in the variable phase space is in equilibrium.

Here, we apply this new theoretical framework to the archetypal, and extremely important, property of fluids: The pressure-volume-temperature ($PVT$) EOS and develop a new generally applicable physics-based (GAP) equation of state for liquids, the GAP equation. We use the term "first principles" to refer to application of relations analytically solved from liquid theory, rather than via computer simulations. General applicability here means that it (a) applies to all liquids regardless of their structure and interactions over a wide $PT$ range (if liquid structure, bonding or conducting type changes as a result of pressure or temperature, the EOS applies to each phase separately) and (b) its derivation is general-theoretical and does not involve any assumptions other than the knowledge of liquid viscosity, and (c) its practical use similarly does not require any additional assumptions other than liquid viscosity. The GAP equation is explicit in the internal energy $E$, and thus leads on naturally from our recent work based on the phonon theory of liquid thermodynamics \cite{ropp,mybook,proctor1,proctor2}. We will expand on the point of general applicability in Section \ref{generality}.

Our GAP equation has only one dimensionless parameter: the Grüneisen parameter for the fluid. Via the dependence on the internal energy, our GAP equation links the most fundamental macroscopic static property of fluids (the pressure-volume-temperature EOS) to their key microscopic property: the molecular hopping frequency or liquid relaxation time. We test our GAP equation against available experimental data for noble Ar and molecular N$_2$ in several different ways and find very good agreement between the GAP equation and experimental data. We observe that the similarity between the solid and liquid equations of state is ultimately related to the condensed state of these phases, whereas the gas EOS is fundamentally different because gas particles are virtually unaffected by cohesion. Finally, we identify the key gaps in the experimental data that need to be filled in to proceed further with the theoretical description of fluids properties from first principles.

\section{Theory}

\subsection{$PVT$ and $PVE$ EOS for solids}
\label{eossol}

There are several forms of the $PVT$ and $PVE$ EOS developed for solids \cite{anderson1}. They all start with the phonon free energy and differentiate it with respect to volume to obtain thermal pressure. Different $PVT$ EOS are related to different approximations and assumptions about the phonon, elastic and thermal properties of solids in order to bridge the gap between theory and experiment. We give an example of a simple EOS for solids and write the high-temperature Helmholtz free energy $F$ as \cite{landaustat}:

\begin{equation}
F=E_0+T\sum\limits_i\ln\frac{\hbar\omega_i}{T}
\label{frees}
\end{equation}

\noindent where $E_0$ is the zero-point energy and $\omega_i$ are the phonon frequencies.

$F$ in \eqref{frees} contributes the thermal pressure $P_{th}=-\left(\frac{\partial F}{\partial V}\right)_T$ to the total pressure $P$:

\begin{equation}
P(V,T)=P_{T=0}(V)+P_{th}(V,T)
\end{equation}

\noindent where $P_{T=0}(V)$ is pressure in the absence of thermal effects \cite{anderson}.

From Eq. \eqref{frees}, $P_{th}$ is

\begin{equation}
P_{th}=\frac{T}{V}\sum\limits_i\gamma_i
\label{free1s}
\end{equation}

\noindent where $\gamma_i=-\frac{V}{\omega_i}\frac{d\omega_i}{dV}$ are mode Gr\"{u}neisen parameters and we neglected the small zero-point term at high temperature. In the high-temperature classical regime, $\gamma_i$ can be set to the average Gr\"{u}neisen parameter $\gamma$: $\gamma_i=\gamma$ \cite{anderson1}.

In solids, the sum in Eq. \eqref{frees} and Eq. \eqref{free1s} is over all $3N$ phonons, where $N$ is the number of particles in the system. This gives

\begin{equation}
P_{th}^s=E_{th}^s\frac{\gamma}{V}=3NT\frac{\gamma}{V}
\label{free4}
\end{equation}

\noindent where $P_{th}^s$ is thermal pressure in the solid and $E_{th}^s$ is thermal energy of the solid.

The total internal energy is the sum of the ``static'' elastic energy at zero temperature, $E_{st}$, and $E_{th}$:

\begin{equation}
E=E_{st}+E_{th}
\end{equation}

Applying $P=-\left(\frac{\partial F}{\partial V}\right)_T$ to the sum gives the Mie-Gr\"{u}neisen (M-G) EOS for high-temperature solids\cite{anderson1}.  Provided that the the function $E_{ST}(V)$ describing the static bond energy dependence on volume is known, the M-G EOS can be written as a $PVE$ EOS

\begin{equation}
PV=-V\frac{dE_{st}}{dV}+\gamma E_{TH}
\label{miegPVE}
\end{equation}

\noindent or, making use of $E_{TH}=3NT$ as in Eq. \eqref{free4} a $PVT$ EOS:

\begin{equation}
PV=-V\frac{dE_{st}}{dV}+3NT\gamma
\label{mieg}
\end{equation}

Differently from some other EOS for solids, the M-G EOS does not contain freely adjustable parameters (fudge factors), for the important reason that it is physics-based. In particular, it is grounded in the statistical theory of the solid state. It is therefore widely applicable to crystalline solids.  The application of the M-G EOS to glasses may be limited by the fact that (since a glass is a metastable state rather than the stable state) the volume of a glass depends on it's history (e.g. cooling rate) \cite{ediger}.  However, the present work deals with liquids, in regions of the phase diagram in which the liquid is the stable equilibrium state.  In the next sections, we will therefore endeavour to derive a similarly physics-based EOS for liquids.

The first term in Eqs. \eqref{miegPVE} and \eqref{mieg} quantifies the static cohesive energy of the solid. This is a system-specific property governed by particular structure, composition and type of interatomic interaction. Therefore, it is not amenable to a general theory (but can be calculated in quantum-mechanical computer simulations). This is in contrast to the second term in Eq. \eqref{mieg} describing thermal properties of solids for which a general theory exists \cite{landaustat}. The way these equations are tested experimentally involves considering the isochores where only the second thermal term contributes to the pressure change. This gives good agreement with experimental EOS measured in many types in solids, including insulators and conductors \cite{anderson1}.

\subsection{Liquid excitations and energy}
\label{liquidexc}

We now set the stage for calculating the EoS for liquids. This calculation is based on the phonon theory of liquid thermodynamics \cite{proctor1,proctor2,prb1,ropp,mybook}. In this section, we recall the starting point and main steps and ingredients of this theory. This is needed in order to follow the physical basis of the liquid EOS.

This theory zeroes in on the liquid internal energy $E$ as the primary property in statistical physics \cite{landaustat} and is based on the fundamental insight from statistical physics that properties of an interacting system is governed by its excitations \cite{landaustat,landaustat1}. In solids, these are collective excitations, phonons. If a solid is amorphous, disorder effects may affect the phonon propagation in a number of ways, yet phonons remain dominant excitations in these systems \cite{zaccone-review,zaccone-book}.

In liquids, collective modes, phonons, include one longitudinal mode and two transverse modes propagating at frequency $\omega>\omega_{\rm F}=\frac{1}{\tau}$ in the solid-like elastic regime \cite{frenkel}. Here, $\tau$ is liquid relaxation time, the time between consecutive particle jumps in the liquid and is related to viscosity $\eta$ as $\tau=\frac{\eta}{G}$, where $G$ is the high-frequency shear modulus \cite{frenkel}. Frenkel arrived at this result by observing that, approximately speaking, the liquid structure does not change on timescales shorter than $\tau$. This implies that at frequency $\omega>\omega_{\rm F}=\frac{1}{\tau}$, liquid supports two solidlike transverse modes.

Experimentally, ascertaining the existence of phonons and their operation in liquids has quite a long history. The last two decades have benefited from powerful synchrotron facilities where inelastic scattering experiments have been performed on many liquids. It is now well established that liquids sustain propagating phonons, both longitudinal and transverse, extending to wavelengths comparable to interatomic separations as in solids \cite{copley,pilgrim,burkel,pilgrim2,water,water-tran,hoso,hoso3,monaco1,monaco2,sn}. This is a remarkable fact asserting close similarity of collective excitations in liquids and solids.

The result for propagating transverse phonons above frequency $\omega_{\rm F}$ can be put on a firm theoretical basis by considering the shear velocity field $v$ derived in the Maxwell-Frenkel viscoelastic theory. This theory yields \cite{ropp,gapreview,mybook}:

\begin{equation}
v\propto\exp\left(-\frac{t}{2\tau}\right)\exp\left(i(kx-\omega t)\right)
\label{gms4}
\end{equation}

\noindent with

\begin{equation}
\omega=\sqrt{c^2k^2-\frac{1}{4\tau^2}}
\label{gms5}
\end{equation}

\noindent where $c$ is the high-frequency transverse speed of sound, $k$ is the wavevector and $\omega$ is the frequency of transverse waves.

According to Eq. \eqref{gms5}, transverse waves exist in liquids only if

\begin{equation}
k>k_g=\frac{1}{2c\tau}
\label{kgap}
\end{equation}

Eq. \eqref{kgap} defines a gapped momentum state present in liquids as well as other dissipative systems \cite{gapreview}. This includes models used to understand the rigidity transition in glasses and liquid-glass transition \cite{naumis}. Here, the $k$-gap increases at low packing fractions. Because rigidity is related to propagating shear modes, the value of $k_g$ was proposed as an order parameter quantifying the rigidity transition. In liquids, the gapped momentum state is consistent with extensive molecular dynamics simulations \cite{yangprl} and predicts that liquids are able to support shear stress at low frequency if system sizes are small enough, the effect ascertained experimentally \cite{noirez1,noirez2}.

According to Eq. (\ref{gms4}), the decay time and decay rate are $2\tau$ and $\Gamma=\frac{1}{2\tau}$. The crossover between propagating and non-propagating modes is usually set by $\omega=\Gamma$. Then, the propagating regime $\omega>\Gamma$ gives $k>\frac{1}{c\tau\sqrt{2}}=k_g\sqrt{2}$ or, using Eq. \eqref{gms5},

\begin{equation}
\omega>\frac{1}{2\tau}
\label{omprop}
\end{equation}

\noindent in agreement with the frequency (energy) gap envisaged originally by Frenkel \cite{frenkel}.

Eqs. \eqref{kgap} and \eqref{omprop} give an important effect mentioned in the Introduction. As $\tau$ decreases at high temperature, both $k_g$ for phonons in Eq. \eqref{kgap} and the threshold frequency for propagating phonons in Eq. \eqref{omprop} increase. This means that the phase space available for these phonons decreases with temperature (and increases with pressure because pressure generally increases $\tau$). Theferore, the phase space available to phonons in liquids is not fixed as in solids but is instead {\it variable} \cite{ropp}.

As long as the system is below the Frenkel line, the longitudinal waves remain propagating in this picture with the usual dispersion relation $\omega=ck$, albeit with different dissipation trends in the regimes $\omega\tau<1$ and $\omega\tau>1$ \cite{ropp}. The longitudinal mode becomes gapped above the Frenkel line as discussed in Section \ref{gaslike}.

The variability of the phase space is a non-perturbative effect itself, and its derivation does not involve a small parameter (recall our discussion in the Introduction that the absence of a small parameter was viewed as a fundamental reason ruling out a general liquid theory). Instead, this variability follows from the Maxwell-Frenkel approach which treats solidlike elastic and hydrodynamic properties of liquids on equal footing and without considering that either component is small \cite{lagrangian,mybook}. However, considering the phonon phase space in liquids and its variability addresses the no small parameter problem stated by Landau, Lifshitz and Pitaevskii. The small parameter does exist in liquids and is the same as in solids: small phonon displacements. Differently from solids where the phase space is due to the fixed number of phonons, $3N$, the small parameter in liquids operates in a reduced phase space where the number of propagating phonons decreases with temperature.

We therefore see that the phonon states in liquids include one longitudinal mode, and two transverse modes with frequency $\omega>\omega_{\rm F}$. The energy of these collective excitations can be calculated using the quadratic Debye density of states \cite{ropp1}. In addition to these excitations, liquids have another type of excitations: local particle jumps enabling liquid flow and setting liquid viscosity. Adding the energy of these excitations to the phonon energy gives \cite{prb1,ropp,mybook,proctor1,proctor2}:

\begin{equation}
E_{th}=NT\left(3-\left(\frac{\omega_{\rm F}}{\omega_{\rm D}}\right)^3\right)
\label{harmo}
\end{equation}

Eq. \eqref{harmo} assumes that the energy of each contributing phonon is given by $T$ in the harmonic approximation.
Accounting for phonon anharmonicity in the Gr{\"u}neisen approximation results in the multiplication of $E_{th}$ by $1+\frac{\alpha T}{2}$, where $\alpha$ is the coefficient of thermal expansion and does not change the energy or heat capacity substantially \cite{ropp,proctor1}.

We note that localised particle jumps give rise to the configurational (communal) entropy of the liquid, $S_c$ \cite{ziman}. In most liquids, $S_c$ is a fairly slowly-varying function of temperature, resulting in a small contribution to the heat capacity $C_v=T\left(\frac{\partial{S_c}}{\partial{dT}}\right)_v$ which can be ignored. In highly anomalous liquids such as water and other tetrahedral systems, $S_c$ may depend on temperature in a wide temperature range due to a continuous coordination change \cite{eisenberg}. We don't consider these system-specific anomalies and instead focus on generic behavior. We also note that water anomalies disappear at high pressures on the order of GPa where water becomes simpler and Ar-like \cite{nist}. Hence our GAP equation becomes applicable to anomalous systems too, albeit at high pressure.

As the hopping frequency $\omega_{\rm F}$ increases, the energy \eqref{harmo} progressively changes from $3NT$ to $2NT$ at high temperature where $\omega_{\rm F}=\omega_{\rm D}$ at the Frenkel line \cite{flreview}, resulting in the decrease of liquid $c_v$ from about $3$ to $2$. This is a universal behavior seen in all liquids \cite{ropp,mybook,proctor1,proctor2,nist}.

Eq. \eqref{harmo} and its extensions to account for anharmonic and quantum effects have been found in agreement with a wide range of liquids in a wide range of temperature and pressures \cite{ropp}. This involves calculating the hopping frequency $\omega_{\rm F}$ from the experimental viscosity $\eta$ as $\omega_{\rm F}=\frac{G}{\eta}$, where $G$ is the high-frequency shear modulus, using this this $\omega_{\rm F}$ in Eq. \eqref{harmo} (or it's quantum equivalent) and calculating $E_{TH}$.  The agreement with experimental data was found \cite{ropp} and subsequently verified independently \cite{proctor1,proctor2}.  The following points from Refs. \cite{ropp,proctor1,proctor2} are relevant to the present work:
\par (1) The parameter that we were calculating was $E_{TH}$, and the testing we performed was comparing $E_{TH}$ to the internal energy output from the fundamental EOS along isochores.  Since the internal energy output from the fundamental EOS is the change in internal energy compared to it's value at some arbitrary point on the phase diagram (rather than it's absolute value), and $E_{ST}$ remains constant along an isochore, it was not necessary to calculate it. The fact that this is the case along isochores we studied is encouraging evidence that assuming $E_{ST}$ is independent of temperature is a good approximation for liquids as well as solids.
\par (2) We found that to reproduce the internal energy output from the fundamental EOS to within ca. 1.5\% it was not necessary to employ any fudge factors.  The only fitting parameters required were two dimensioned, and physically meaningful quantities.  To achieve the more challenging task of reproducing the observed heat capacities accurately it was necessary to allow the liquid relaxation time to vary by ca. 2\% from the value obtained from the viscosity.  However, using the model developed in Ref. \cite{ropp,proctor1,proctor2} to provide input data for a PVE EOS only requires the internal energy, not the heat capacity.
\par (3) Our model provided the thermal energy $E_{TH}$ as a function of either the viscosity or the liquid relaxation time.  So to place this in the context of the present work we can say that we developed an $E_{TH}\eta$ EOS and an $E_{TH}\omega_{\rm F} (E_{TH}\tau)$ EOS for liquids.

\subsection{$PVT$ and $PVE$ GAP equations of state for liquids}
\label{oureos}

\subsubsection{Free energy}

We now use the results from the previous section \ref{liquidexc} to derive the EOS for liquids. The thermal part of the Helmholtz free energy, $F$, can be evaluated using the relation:

\begin{equation}
E_{th}=-T^2\left(\frac{\partial}{\partial T}\frac{F}{T}\right)_V
\label{efromf}
\end{equation}

Using $E_{TH}$ in Eq. \eqref{harmo} and integrating the energy implies the following terms in $F$:

\begin{equation}
F\propto-3NT\ln(T)+NT\int\left(\frac{\omega_{\rm F}}{\omega_{\rm D}}\right)^3\frac{dT}{T}+gT
\label{free}
\end{equation}

\noindent where $g$ is the function which is related to the integration constant and which does not depend on temperature.

$g$ should include the term $g_1$ such that when the hopping frequency $\omega_{\rm F}=0$, the free energy is equal to the high-temperature free energy of the solid $F_s$ as $F_s=3NT\ln(\hbar\omega)-3NT\ln(T)$, where $\omega$ is the average geometric phonon frequency \cite{landaustat}. This is supported by the close similarity of phonon states in solid and liquid states \cite{copley,pilgrim,burkel,pilgrim2,water,water-tran,hoso,hoso3,monaco1,monaco2,sn,ropp} and, therefore, similarity of the corresponding free energies. This gives $g_1=3N\ln(\hbar\omega)$ and implies the following terms in the free energy

\begin{equation}
F\propto 3NT\ln\frac{\hbar\omega}{T}+NT\int\left(\frac{\omega_{\rm F}}{\omega_{\rm D}}\right)^3\frac{dT}{T}
\label{free1}
\end{equation}

Similarly to $g_1$ setting the first term in Eq. \eqref{free1}, another term in $g$ should complement the second term to give it the dimensions of energy. Regardless of what this term is, we see that a general form of $F$ can be written as

\begin{equation}
F=3NT\ln\frac{\hbar\omega}{T}+f\left(NT\int\left(\frac{\omega_{\rm F}}{\omega_{\rm D}}\right)^3\frac{dT}{T}\right)
\label{free2}
\end{equation}

\noindent where $f$ is a smooth function satisfying $f=0$ when $\omega_{\rm F}=0$ as discussed above.

The first term in Eq. \eqref{free2} is the free energy of the solid. The second term is our liquid term due to finite viscosity $\eta=\frac{G}{\omega_{\rm F}}$. $F$ becomes the free energy of the solid when $\omega_{\rm F}=0$ as expected.

The exact form of $f$ is unimportant for the EOS we are going to derive in the next section because we will consider the limit $\frac{\omega_{\rm F}}{\omega_{\rm D}}\ll 1$ at which the second term in Eq. \eqref{free2} becomes small. However, we note that setting $f(x)=x$
and applying Eq. \eqref{efromf} to Eq. \eqref{free2} gives $E_{th}$ equal to the liquid energy in Eq. \eqref{harmo}.


\subsubsection{Two forms of the GAP equation}
\label{twoforms}

The $PVT$ EOS follows from applying $P_{th}=-\left(\frac{\partial F}{\partial V}\right)_T$ to Eq. \eqref{free2}. In the first term of Eq. \eqref{free2}, the geometric mean frequency $\omega$ depends on $V$ and results in the Mie-Gr\"{u}neisen (M-G) EOS for solids. In the second term, the volume dependence is largely contained in $\omega_{\rm F}$ set by viscosity. $\omega_{\rm D}$ depends on $V$ too, although to a smaller degree.

This gives the first form of the GAP equation as

\begin{equation}
P_{th}=\frac{3NT\gamma}{V}-NTf^\prime\frac{\partial}{\partial V}\int\left(\frac{\omega_{\rm F}}{\omega_{\rm D}}\right)^3\frac{dT}{T}
\label{eos1}
\end{equation}

\noindent where $f^\prime$ is the derivative of $f$.

In the low-temperature viscous regime $\omega_{\rm F}\ll\omega_{\rm D}$, the second term is small and can be ignored. The important point is that this applies to the liquid state a wide range of the phase diagram.

Let us first consider viscous liquids \cite{dyre,mybook} and recall that $\omega_{\rm F}=\frac{G}{\eta}$ implies $\frac{\omega_{\rm F}}{\omega_{\rm D}}\approx\frac{\eta_0}{\eta}$, where $\eta_0$ is the high-temperature limiting value of viscosity. In viscous melts, this ratio is extremely small. For SiO$_2$, a common example as well as very common component of many viscous melts, $\eta$ at melting temperature is approximately 10$^6$ Pa$\cdot$ s \cite{ojovan}. Typical values of the limiting viscosity $\eta_0$ are 10$^{-5}$-10$^{-4}$ Pa$\cdot$ s \cite{nussinov,sciadv}. This is close to $\eta$ of about 10$^{-3}$ Pa$\cdot$s in simulated SiO$_2$ where $\eta$ saturates to a constant \cite{horbach}. This gives $\frac{\omega_{\rm F}}{\omega_{\rm D}}\approx\frac{\eta_0}{\eta}$ of the order of $10^{-9}$ and similarly small values in other viscous liquids and melts. This ratio enters as a cube in Eq. \eqref{eos1}).

$\frac{\omega_{\rm F}}{\omega_{\rm D}}\ll 1$ also applies to low-viscous liquids such as water. Note that water viscosity at room temperature is interestingly not far above the minimal quantum viscosity \cite{pt2021,myreview}, the lowest viscosity that any liquid can ever attain \cite{sciadv}. Yet even in this case, $\frac{\omega_{\rm F}}{\omega_{\rm D}}$ is small. Let us consider two temperatures: melting temperature $T_m$ and room temperature $T_r$. Experimentally, $\frac{\eta_0}{\eta(T=T_m)}$ and $\frac{\eta_r}{\eta(T=T_r)}$ are 0.159 and 0.346 \cite{nist}, implying the same values for $\frac{\omega_{\rm F}}{\omega_{\rm D}}$. This is consistent with X-ray scattering experiments showing the viscoelastic behavior of water where molecules undergo many oscillations before jumping to new quasi-equilibrium positions in a wide temperature range of about ($T_m, T_m+100$ K) \cite{water}, implying $\omega_{\rm F}\ll{\omega_{\rm D}}$. $\frac{\omega_{\rm F}}{\omega_{\rm D}}=0.159$ and $\frac{\omega_{\rm F}}{\omega_{\rm D}}=0.346$ give $\left(\frac{\omega_{\rm F}}{\omega_{\rm D}}\right)^3=4\cdot 10^{-3}$ and $\left(\frac{\omega_{\rm F}}{\omega_{\rm D}}\right)^3=4\cdot 10^{-2}$ in Eq. \eqref{eos1}.

Dropping small $\left(\frac{\omega_{\rm F}}{\omega_{\rm D}}\right)^3$ in Eq. \eqref{eos1} simplifies the GAP equation to

\begin{equation}
P_{th}=\frac{3NT\gamma}{V},
\label{eosl}
\end{equation}

\noindent the M-G EOS discussed in Section \ref{eossol}.

The second general form of the GAP equation follows from recalling

\begin{equation}
\omega_{\rm F}=\omega_{\rm D}e^{-\frac{U}{T}}
\label{omegaf}
\end{equation}

\noindent where $U$ is the activation energy for particle jumps from one quasi-equilibrium place to the next \cite{frenkel,dyre}.

$U$ is constant at high temperature in the low-viscous regime. For some (``fragile'') liquids, $U$ can be temperature-dependent, however this becomes a significant effect in the viscous supercooled regime only \cite{dyre}. This temperature dependence of $U$ disappears again at yet lower temperature where $U$ crosses over to a constant value. This corresponds to the crossover from the Vogel-Fulcher-Tammann to Arrhenius dependence of $\tau$ at $\tau$ of about 10$^{-6}$ s at the Stickel crossover \cite{cro1,cro2,cro3,cro4,cro5}.

Using Eq. \eqref{omegaf} in Eq. \eqref{eos1} gives the second general form of the GAP equation as

\begin{equation}
P_{th}=\frac{3NT\gamma}{V}-NTf^\prime\int\frac{\partial}{\partial V}e^{-\frac{3U}{T}}\frac{dT}{T}
\label{eos2}
\end{equation}

In the second term, volume dependence is contained in $U$: the activation energy generally increases with pressure $P$ and decreases with volume ($\frac{\partial U}{\partial V}<0$). This can be seen in more detail as follows. At zero or small pressure, $U$ is set by internal elasticity of the liquid. In particular, $U$ is the energy of elastic deformation required for the atomic cage to increase its size from radius $r$ to $\Delta r$ to enable the central atom to escape the cage. Calculating this energy gives $U=8\pi G \Delta r^2r$, where $G$ is the instantaneous shear modulus \cite{frenkel,dyre}. Applying external pressure $P$ increases $U$ because it adds extra work required to expand the cage, $4\pi r^2\Delta rP$. $U$ becomes

\begin{equation}
U=8\pi G \Delta r^2r+4\pi r^2\Delta rP
\end{equation}

\noindent and increases with pressure (decreases with volume).

Assuming smooth behavior of all functions and weak or no temperature dependence of $U$, taking the volume derivative in Eq. \eqref{eos2} and integrating gives

\begin{equation}
P_{th}=\frac{3NT\gamma}{V}+NTf^\prime\frac{1}{U}\frac{\partial U}{\partial V}e^{-\frac{3U}{T}}
\label{eos3}
\end{equation}

Eq. \eqref{eos3} is a closed-form of the second GAP EOS for liquids. The second term contains both volume dependence and temperature dependence, giving an exponential temperature correction to the first linear term. This correction is negative because $\frac{\partial U}{\partial V}<0$. If $f(x)=x$ as discussed earlier, $f^\prime=1$.

Consistent with what we saw earlier, at low temperature $T\ll U$ (or, equivalently, $\omega_{\rm F}\ll\omega_{\rm D}$, in view of Eq. \eqref{omegaf}) in the viscous regime, the second term in Eq. \eqref{eos3} is small and can be dropped, resulting in Eq. \eqref{eosl}.

In this subsection, we considered the case $\omega_{\rm F}\ll\omega_{\rm D}$ in two different forms of the GAP equations and noted that this applies to a wide range of the liquid phase diagram. We now consider what happens at higher temperature and pressure when the condition $\omega_{\rm F}\ll\omega_{\rm D}$ does not apply. The hopping frequency $\omega_{\rm F}$ increases with temperature exponentially according to Eq. \eqref{omegaf}, whereas the variation of its limiting value, $\omega_{\rm D}$, is slower. In the narrow range where both frequencies become comparable, $\omega_{\rm F}\approx\omega_{\rm D}$, the condition we used earlier, $\omega_{\rm F}\ll\omega_{\rm D}$, no longer applies. As $\omega_{\rm F}$ tends to its limiting value $\omega_{\rm D}$ in the regime $\omega_{\rm F}\approx\omega_{\rm D}$, the volume-dependent ratio $R$

\begin{equation}
R=\frac{\omega_{\rm F}}{\omega_{\rm D}}
\end{equation}

\noindent in the second term in Eq. \eqref{eos1} becomes close to a constant, 1, and a slowly-varying function of $V$. As a result, the volume derivative in the second term of Eq. \eqref{eos1} becomes small and can be dropped. We therefore find that, similarly to the low-temperature case where $R\ll 1$, the EoS in this high-temperature regime is, to a good approximation, given by Eq. \eqref{eosl}.

\subsubsection{Generality of the liquid EOS}
\label{generality}

In the Introduction, we briefly commented on the generality of our EOS for liquids. Now is the good time to expand on this point. We first observe that in addition to the intra-cage rattling, or Debye, frequency $\omega_{\rm D}$, Eq. \eqref{eos1} contains the inter-cage frequency $\omega_{\rm F}$. This frequency quantifies the rate of particle jumps between neighbouring quasi-equilibrium positions in the liquid and is related to liquid viscosity as $\omega_{\rm F}=\frac{G}{\eta}$. The presence of $\omega_{\rm F}$ in the liquid theory is understandable because liquids flow, and $\omega_{\rm F}$ quantifies this flow.

The generality of the EOS \eqref{eos1} comes from several features. First, this equation has no free fitting parameters: $R=\frac{\omega_{\rm F}}{\omega_{\rm D}}$ is fixed by the system properties.

Second, Eq. \eqref{eos1} applies to liquids with different interatomic interactions and correlation functions. As long as $R$ and its derivative are the same for these different liquids, the EOS is predicted to be the same. This provides a wide range of applicability and universality of this EOS, particularly if compared to those approaches to the EOS which rely on system-specific structure and interactions. We recall our discussion in the Introduction related to several important issues faced by the approach to liquids based on interactions and correlation functions. These interactions and functions are generally complex and unknown apart from simple cases such as Lennard-Jones and related systems.

While the issues of complexity and availability are to some extent of a practical character, the other issue is more foundational and is related to fundamental understanding of the liquid state. This issue is related to the observation by Landau, Lifshitz and Pitaevskii discussed in the Introduction: interactions in liquids are strong and system-specific, therefore, liquid thermodynamic properties are also system-specific, precluding the calculation of liquid thermodynamic properties in general form, in contrast to solids and gases. This is the no small parameter problem of liquid theory discussed in the Introduction. The same applies to the $PVT$ or $PVE$ EOS which is obtained from the derivative of the free energy $F$. Here, this problem is overcome by considering excitations in liquids discussed in Section \ref{liquidexc} instead of system-specific interactions and structure.

We therefore see that addressing the no small parameter problem for liquid energy and other thermodynamic functions discussed in Section \eqref{liquidexc} naturally solves the problem of general applicability of the liquid EOS.

We note that the GAP EOS \eqref{eos1} is not as general as the EOS in solids \eqref{free4} and contains the factor $\frac{\omega_{\rm F}}{\omega_{\rm D}}$. Such generality is impossible to achieve in liquids because liquid properties depend on viscosity $\eta=\frac{G}{\omega_{\rm F}}$ which is strongly temperature-dependent. Therefore, $\omega_{\rm F}$ must enter the thermodynamic properties and their derivatives such as the EOS. Nevertheless, our approach to liquids based on excitations goes a long way towards describing the EOS for liquids in much more general terms. This is achieved by getting rid of system-specific interactions and correlation functions operating in terms of one single parameter $\omega_{\rm F}$ instead. This parameter describes the liquid flow. As discussed in Section \ref{liquidexc}, this parameter also governs the phase space available to phonons in liquids.

As a final observation in this section, we recall that the hopping frequency is related to viscosity as $\omega_{\rm F}=\frac{G}{\eta}$.
Viscosity is readily measured regardless of liquid complexity related to structure and interactions. Moreover, for some liquids viscosity and relaxation time are found to scale with volume and temperature \cite{scale1,scale2,scale3}. This aids in developing models of viscosity, extending it to temperature and pressure where no viscosity measurements currently exist and then using the corresponding $\omega_{\rm F}$ in Eq. \eqref{eos1}.

\subsubsection{Other comments related to the GAP equations}

As noted in the previous subsection, Eq. \eqref{eosl} following from $\frac{\omega_{\rm F}}{\omega_{\rm D}}\ll 1$ is a very good approximation to the liquid EOS in a wide range of the phase diagram. Physically, the reason for this is that the primary property, the liquid energy \cite{landaustat}, is to a very good approximation equal to the solid energy $E=3NT$ when $\omega_{\rm F}\ll\omega_{\rm D}$. This is readily seen from Eq. \eqref{harmo}. We will develop this point in the Discussion section \ref{discussion} where we will note that the similarity between liquid and solid properties is ultimately due to the fact that both phases are condensed states. On the other hand, liquids and gases are qualitatively different states of matter because gases are not condensed states. This revisits the point we made in the Introduction: historical use of gas-like models to describe liquid properties understandably gives results inconsistent with experimentals in real liquids.

The finding that the GAP EOS reduces to the simple form Eq. \eqref{eosl} in the wide range of the liquid phase diagram might at first sight be taken as somewhat disappointing from the point of view of advancing a theoretical EOS. We make two points in this regard. First, this simplicity is a natural and a physically necessary result in view of the similarity between important properties of liquids and solids which are both condensed states of matter (we will expand on this point in the Discussion section \ref{discussion}). A different mathematical EOS for liquids would be physically inapplicable.
Second, even though the working approximation for the GAP equation for the liquid phase diagram turns out to be simple, it is the proper derivation of this result which is theoretically important here. This is because deriving this result using a general liquid theory is not at all simple: recall the Introduction discussing long-standing fundamental problems of general theory of liquid thermodynamic properties including the no small parameter problem in a liquid theory. In Section \ref{liquidexc}, we showed how the theory based on excitations in the system enables to overcome these problems and make sense of many hitherto unexplained results. For {\it some} properties such as liquid heat capacity, this theory gives nontrivial results in agreement with experimental data that exhibit different trends to those observed for solids \cite{ropp,mybook,proctor1,proctor2} as discussed in Section \ref{liquidexc}. For other properties such as the liquid EOS considered here, the theory gives results analogous to the outcomes for solids. Both outcomes are equally valuable from the physical point of view. Importantly, this includes the ability of the GAP equation and its approximation to understand liquid experimental data in a wide range of temperature and pressure. We will discuss this in Section \ref{testing}.

The apparent simplicity of the EOS in Eq. \eqref{eosl} can be contrasted to the non-trivial behavior of the liquid thermodynamic functions such as energy and specific heat discussed in the Introduction. The reason for this contrast is as follows. The factor $R=\frac{\omega_{\rm F}}{\omega_{\rm D}}$ is present in energy \eqref{harmo} as well as free energy \eqref{free2}. When $R$ becomes 0 or 1, $E$ in Eq. \eqref{harmo} undergoes significant changes, resulting in the decrease of specific heat from about $3$ at $R=0$ to $2$ at $R=1$ \cite{ropp,proctor1}. On the other hand, when $R$ becomes either 0 or 1, the second term of the free energy in Eq. \eqref{free2} stops being dependent on volume as discussed earlier. As a result, the volume derivative in the second term in Eq. \eqref{eos1} becomes small at both $R=0$ and $R=1$, and Eq. \eqref{eos1} simplifies to Eq. \eqref{eosl} in both cases.

Eqs. \eqref{eos1}, \eqref{eosl}, \eqref{eos2}, \eqref{eos3} describe different forms of thermal pressure contributing to a $PVT$ or $PVE$ EOS for liquids in terms of the GAP equations. These forms are a physics-based model rather than a convenient fitting function with many freely adjustable parameters as was the case with other EOS. We are now able to write the full EOS by adding the static term discussed in Section \ref{eossol}. We saw that Eq. \eqref{eosl} gives a very good approximation to the thermal pressure of liquids in a wide range of pressure and temperature on the phase diagram where $\omega_{\rm F}\ll\omega_{\rm D}$. This gives the approximation to the GAP equation which is similar to the EoS of solids \eqref{mieg} discussed in section \ref{eossol}:

\begin{equation}
PV=-V\frac{dE_{st}}{dV}+3NT\gamma
\label{eosl1}
\end{equation}

\noindent where $E_{st}$ is the zero-temperature static energy due to cohesive forces, the second term is the thermal term and $\gamma$ is the liquid Gr\"{u}neisen parameter.

We also saw that the same condition, $\omega_{\rm F}\ll\omega_{\rm D}$, results in $E_{th}=3NT$ according to Eq. \eqref{harmo}. This enables us to write Eq. \eqref{eosl1} in a more general form involving the thermal liquid energy $E_{th}$ as

\begin{equation}
PV=-V\frac{dE_{st}}{dV}+\gamma E_{th}
\label{eosl2}
\end{equation}

Eqs. \eqref{eosl1} and \eqref{eosl2} give us several ways to experimentally test this equation experimentally. In section \ref{constr}, we will discuss how it can be usefully applied to real liquids and predict their $PV$ EOS.

As a final observation in this section, we recall the Rosenfeld's conjecture \cite{rosenscaling} mentioned in the Introduction. The conjecture proposes that liquid transport properties including viscosity scale with the excess entropy of the liquid, even though there are conceptual problems with clear definitions of this entropy. Eq. \eqref{harmo} and Eqs. \eqref{free}-\eqref{free2} explain why thermodynamic properties including the entropy may correlate with viscosity. According to these equations, liquid energy, heat capacity, free energy and its derivatives such as entropy depend on the hopping frequency $\omega_{\rm F}=\frac{1}{\tau}=\frac{G}{\eta}$. Hence, the entropy as well as other thermodynamic properties such as energy (Eq. \eqref{harmo}) and heat capacity correlate with viscosity $\eta$.

\subsection{The gas-like state of liquid dynamics}
\label{gaslike}

Later on in this paper, we will be comparing our GAP equation derived in the previous section to the experimental data which are generally not available much above 300 K at liquid-like densities. Nevertheless, it is important to discuss how our approach needs to be modified in the high-temperature part of the phase diagram. This is important from the point of view of theory as well as comparing this theory to high-temperature experimental data when it becomes available in the future. It is also useful from the point of view of testing consistency of the theory and its ability to give sensible results in different parts of the phase diagram.

In the previous section, we discussed the liquidlike particle dynamics where each particle undergoes many inter-cage oscillations with frequency $\omega_{\rm D}$ or period $\tau_{\rm D}=\frac{1}{\omega_{\rm D}}$ and jumps to the next quasi-equilibrium state with inter-hopping frequency $\omega_{\rm F}=\frac{1}{\tau}$, where $\tau$ is liquid relaxation time. When $\tau\approx\tau_{\rm D}$, the oscillatory component of particle motion is lost and the diffusive gas-like hopping component remains. This corresponds to the crossover at the Frenkel line mostly lying above the critical point \cite{flreview} and gives liquid specific heat of about $c_v=2$ \cite{ropp}. We know that $c_v$ should tend to its gas-like value of $\frac{3}{2}$ at high temperature. This can happen either abruptly when the vapour pressure curve is crossed or gradually if the system is above the critical point. We now consider the EOS in this high-temperature/low pressure part of the phase diagram where particle dynamics qualitatively changes and becomes gas-like.

When $\tau\approx\tau_{\rm D}$, the $k$-gap in Eq. \eqref{kgap} approaches the zone boundary ($c\tau_{\rm D}\approx a$, where $a$ is the interatomic separation), and all transverse modes disappear. The remaining excitation in the system is the longitudinal wave. As a result, $c_v$ undergoes a crossover because the temperature dependence of energy changes \cite{lingcross}. Hence instead of Eq. \eqref{harmo} describing the disappearance of two transverse waves, we need to consider the evolution of the remaining longitudinal wave.

In the gaslike regime of liquid dynamics, we use the gas kinetic theory and the particle mean free path $L$. The wavelengths at which the system can oscillate are larger than $L$ because at shorter distances the dynamics is ballistic rather than oscillatory. Then, the liquid energy is (compare it to Eq. \eqref{harmo}) \cite{ropp,mybook}:

\begin{equation}
E_{th}=\frac{3}{2}NT+\frac{1}{2}NT\left(\frac{a}{L}\right)^3
\label{super}
\end{equation}

As expected, the energy in the gas-like state \eqref{super} at $a=L$, $E=2NT$, matches the energy in the liquidlike state \eqref{harmo} when $\omega_{\rm F}=\omega_{\rm D}$ at the Frenkel line. The corresponding $c_v$ is close to 2, serving as a thermodynamic definition of the Frenkel line \cite{ropp,flreview}.

$L$ can be evaluated from experimental gas-like viscosity as $\eta=\frac{1}{3}\rho vL$, where $\rho$ is density and $v$ is average thermal velocity. This gives good agreement with experimental $c_v$ for several supercritical systems, including noble and molecular fluids \cite{ropp}.

The EOS of liquids in the gas-like regime can be discussed in terms similar to those in the previous section. The free energy corresponding to the first term in Eq. \eqref{super} is the free energy of the ideal gas \cite{landaustat}:

\begin{equation}
F_{id}=-NT\ln\frac{eV}{N}+Nh(T)
\label{freeid}
\end{equation}

\noindent where $h(T)$ depends on temperature only.

The free energy corresponding to the non-ideal second term in Eq. \eqref{super} contains the term $\left(\frac{a}{L}\right)^3$, similarly to the free energy in the liquidlike regime of dynamics containing $\left(\frac{\omega_{\rm F}}{\omega_{\rm D}}\right)^3$ in Eq. \eqref{free2}. This gives the total free energy as

\begin{equation}
F=-NT\ln\frac{eV}{N}+Nh(T)-g\left(\frac{NT}{2}\int\left(\frac{a}{L}\right)^3\frac{dT}{T}\right),
\label{freeabovefl}
\end{equation}

\noindent where $g$ is a smooth function satisfying $g(0)=0$, and the thermal pressure $P_{th}=-\left(\frac{\partial F}{\partial V}\right)_T$ as

\begin{equation}
P_{th}=\frac{NT}{V}+\frac{NT}{2}g^\prime\frac{\partial}{\partial V}\int\left(\frac{a}{L}\right)^3\frac{dT}{T}
\label{pabovefl}
\end{equation}

As expected, this equation gives the ideal-gas EOS at high temperature. Indeed, the volume dependence of the thermal pressure in the second term is contained in the mean free path $L$ through viscosity and interatomic separation $a$. When $L$ exceeds $a$, the last term in Eq. \eqref{freeabovefl} is small and can be ignored (in practice, this becomes a good approximation in a fairly dense state where $L\approx 2a$ already due to the third power of $\frac{a}{L}$ in Eq. \eqref{pabovefl}.
As a result, $PV=NT$ is a good approximation to the fluid EOS in the range of the phase diagram where particle dynamics is gas-like above the Frenkel line \cite{flreview}.

\subsection{Constructing equations for $E_{ST}(V)$}
\label{constr}

To apply our new equation we need to construct an expression for $E_{ST}(V)$. A model-free way to treat this term is to do the same testing as in the case of solids where the M-G EOS is tested on isochores. In that case, the static term does not contribute and $PV$ in Eq. \eqref{mieg} is proportional to temperature with the slope set by the Gr\"{u}neisen parameter \cite{anderson}. As discussed in Section \ref{twoforms}, the GAP equation for liquids is approximated by the M-G EOS in a wide range on the phase diagram so this method of testing is at the same level as in the solid state theory.

In developing $PV$, $PVT$ and $PVE$ EOS, models for $E_{ST}(V)$ need to be introduced which involves some departure from first principles. This is the case for developing both the EOS for solids \cite{anderson} and liquids. In the present work we are going to consider liquids at sufficient density for the repulsive part of $E_{ST}(V)$ to dominate.  The strong repulsion preventing atoms from overlapping originates from (a) electrostatic repulsion between the electron clouds and (b) the Pauli exclusion principle.  There is therefore no succinct analytical expression available backed by first principles.  In the present work we will compare three different approaches.
The first approach is derived from the repulsive part of the Lennard-Jones (LJ) potential.  The Lennard-Jones potential is frequently used without question as if it comes from first principles, however this is not in fact the case (\cite{PruteanuLJ} and refs therein).  Nonetheless, the LJ potential has been used very successfully for many applications relating to fluids for decades.  It is therefore a reasonable option for the construction of $E_{ST}(V)$.  Writing the $1/r^{12}$ law in terms of volume, we obtain the following equation for $E_{ST}(V)$:

\begin{equation}
E_{ST}(V) = A/V^{4}
\label{LJrepulsion}
\end{equation}

For the other two approaches, we will relate $E_{ST}(V)$ to the static component of the pressure via $E_{ST}(V)=P_{ST}V$.  The second approach is to assume a constant bulk modulus:

\begin{equation}
B_{ST} = -V\frac{dP_{ST}}{dV}
\label{ConstantB1}
\end{equation}

Integrating leads to:

\begin{equation}
P_{ST} = -B_{ST}\ln{\frac{V}{V_0}}
\label{ConstantB2}
\end{equation}

\noindent and

\begin{equation}
E_{ST} = -B_{ST}V\ln{\frac{V}{V_0}}
\label{ConstantB3}
\end{equation}

Here, $V_0$ is the (hypothetical) volume at $P_{ST}=0$.

The third approach is that most commonly used for the solid state, which is to assume a linearly pressure dependent bulk modulus $B_{ST}=B_0+B'P_{ST}$.

In this case the following equation for $E_{ST}$ is obtained:

\begin{equation}
E_{ST} = \frac{B_0}{B'}\left[ \frac{V^{1-B'}}{V_0^{-B'}} - 1 \right]
\label{BPrime}
\end{equation}

In each case, $\frac{dE_{ST}}{dV}$ can be obtained analytically from the relevant equation for $E_{ST}$.

\section{Comparing to experimental data}
\label{testing}

\subsection{Preliminary notes}

In this section we test our new GAP equation against available experimental data in the form of a $PVE_{TH}$ EOS (equation \eqref{eosl2}) and a $PVT$ EOS (equation \eqref{eosl1}).  For the testing in $PVE_{TH}$ form, $E_{TH}$ is derived from experimentally measured quantities as follows.  The total internal energy of the liquid $E$ is related to the static energy (which we calculate anyway in order to obtain $\frac{dE_{st}}{dV}$) and the thermal energy according to $E=E_{th}+E_{st}$.  Internal energy outputs closely matching the available experimental data (subject to constraints below) are available from the fundamental EOS for a variety of simple fluids over a reasonable $P,T$ range.  However, these data are measured relative to some arbitrary reference point $E_0$ (in any case the absolute value of internal energy cannot, in a strict sense, be measured experimentally unless one heats up the sample from absolute zero).  We therefore denote the internal energy output from the fundamental EOS by $E^\prime$, and write $E=E^\prime-E_0$.  Combining these, we can obtain the thermal energy as follows:

\begin{equation}
E_{th}=E^\prime-E_{st}-E_0
\end{equation}

However, before using these data it is important to check the extent to which the internal energy output from the fundamental EOS is backed by experimental data. Upon temperature change along an isochore, the change in internal energy can be calculated from the heat capacity data and is positive for fluids under all conditions:

\begin{equation}
\delta E=C_v\delta T
\label{j1}
\end{equation}

Upon pressure change along an isotherm the change in internal energy depends on the $PV$ EOS and the thermal expansion coefficient. It is given by the following equation \cite{sears}:

\begin{equation}
\delta E=-\Bigl[T\left(\frac{\partial V}{\partial T}\right)_P+P\left(\frac{\partial V}{\partial P}\right)_T\Bigr]\delta P
\label{j2}
\end{equation}

This equation is valid for fluids under all conditions but results in strikingly different behaviour for an ideal gas ($\delta E=0$), a real fluid inside the Amagat curve ($\left(\frac{\partial E}{\partial P}\right)_T<0$) and a real fluid outside the Amagat curve ($\left(\frac{\partial E}{\partial P}\right)_T>0$). Our models for $E_{st}(V)$ consider only repulsion between fluid particles. In this case, our GAP equation applies when we are both outside the Amagat curve and (as discussed earlier) on the liquid-like side of the Frenkel line \cite{flreview}.

From Eqs. \eqref{j1} and \eqref{j2} it is clear that for the internal energy output from the fundamental EOS to be reliable it should be backed by both $PVT$ EoS data and heat capacity data.  Whilst direct measurement of heat capacities at extreme pressures and temperatures are challenging, the heat capacities are constrained by their relation to the speed of sound.  Speed of sound measurements are typically available over a wider $P,T$ range than the heat capacity data.  Even with this in mind, it is only for Ar \cite{tegelerArEOS} and N$_2$ \cite{spanN2EOS} to our knowledge that adequate data exist at $P,T$ conditions outside the Amagat curve and on the liquid-like side of the Frenkel line to conduct meaningful testing of our new GAP equation. For this reason, we consider the experimental data on Ar and N$_2$ below and compare this data to the GAP equation.

\subsection{Testing the GAP EOS against experimental data for Ar}

\begin{figure*}
{\scalebox{0.35}{\includegraphics{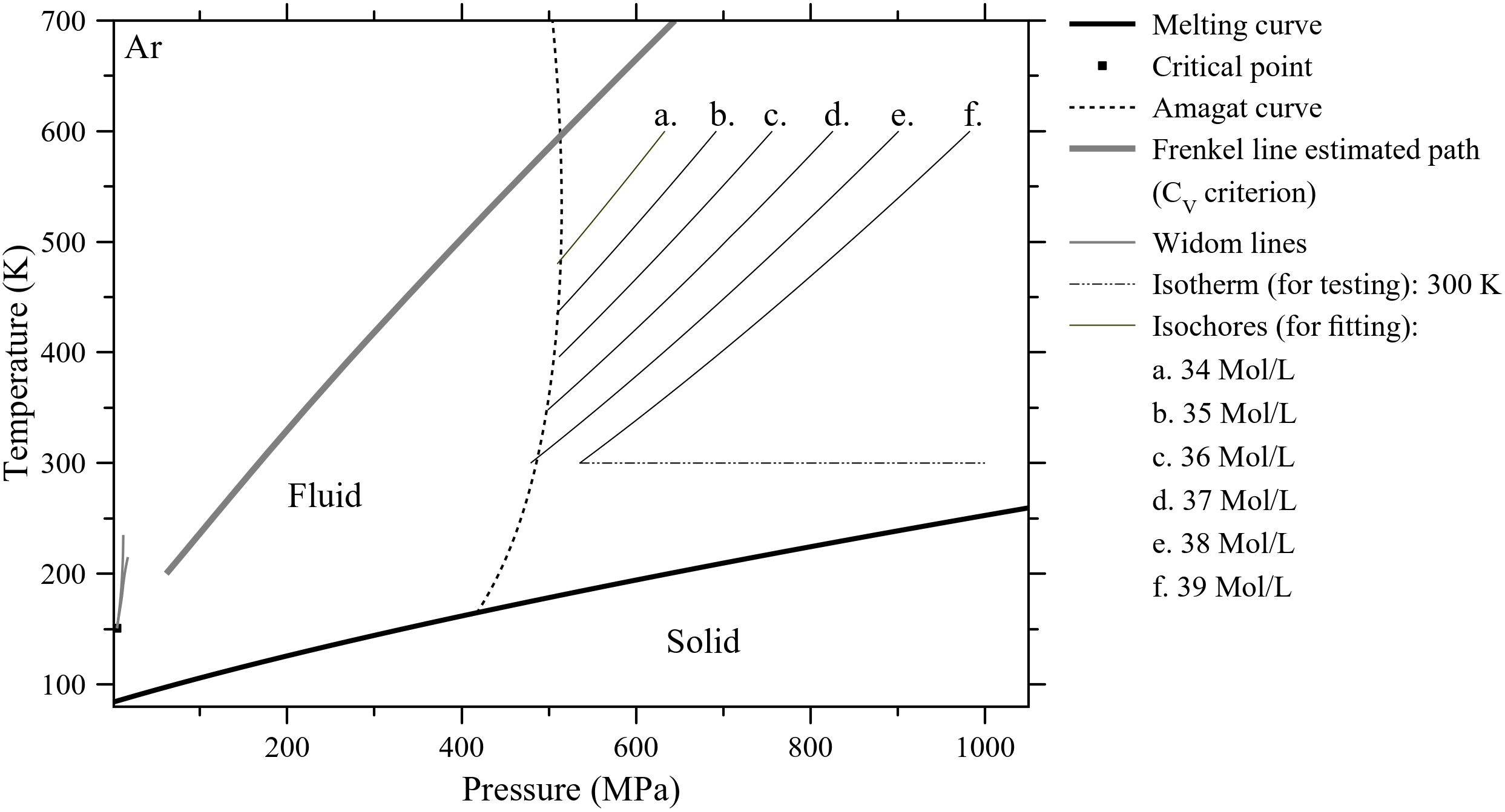}}}
\caption{Pressure-temperature phase diagram of Ar (produced using the methodology outlined in Ref. \cite{proctor2}). $P$,$T$ paths along which our GAP equation has been fitted and tested are marked.}
\label{fig1}
\end{figure*}

Figure \ref{fig1} shows the $P,T$ phase diagram of Ar. The isochores and isotherm along which testing has taken place are marked.

Substituting for $E_{th}$ in equation \eqref{eosl2} results in the EOS written as follows:

\begin{equation}
PV=-V\frac{dE_{st}}{dV}+\gamma[E^\prime(V,T)-E_{st}(V)-E_0]
\label{j3}
\end{equation}

This equation therefore has (unavoidably) 2-4 adjustable parameters. However, it is important to note that these are not freely adjustable parameters (``fudge factors'') but are fixed by system properties. Using this kind of parameter is common in physics \cite{myreview} and in fact is what physics is considered to be about: one view holds that the essence of every physical theory is to predict a future experiment on the basis of a previous one \cite{landaupeierls} or, in other words, provide a relationship between different properties of the system.
The first adjustable parameter is $\gamma$ and the rest depend on the approach taken for $E_{st}(V)$. The Lennard-Jones approach has a single parameter ($A$), the constant bulk modulus approach has two parameters ($B_{ST}$ and the constant volume $V_0$ arising when $E_{st}$ is obtained by integrating), whilst the pressure-dependent bulk modulus approach has these parameters as well as $B^\prime$, the derivative of the bulk modulus with respect to pressure.

The most rigorous approach to finding the values for these parameters is to use regression analysis, and make as much as possible of the regression analysis linear.  Beginning by fitting to the data along isochores makes this possible. If we plot $PV$ against $E^\prime (V,T)$ along an isochore the gradient should be linear according to equation \eqref{eosl2}, written as $PV=i(V)+\gamma E^\prime (V,T)$ where:

\begin{equation}
i(V)=-V\frac{dE_{st}}{dV}+\gamma[-E_{st}(V)-E_0]
\label{j31}
\end{equation}

The obtained values of $\gamma$ and $i(V)$ for the various Ar isochores from figure \ref{fig1} are given in Table 1. Importantly, we observe that the obtained values of $\gamma$ are in the physically-sensible range 1.7-2. This is consistent with the range of the Gr\"{u}neisen parameters commonly observed in condensed matter including solids \cite{anderson1}. This is an important test of our theory: if the fitting resulted in the unphysical Gr\"{u}neisen parameters, the theory would have to be reconsidered.

\begin{table}
\begin{tabular}{l c l c l c}
\hline
Density (Mol/m$^3$) & $\gamma$ & i(V) (J)\\
\hline
34000 & 1.745(2) & 10212(9)&\\
35000 & 1.795(1) & 11069(4)&\\
36000 & 1.844(5) & 11978(19)&\\
37000 & 1.898(7) & 12924(21)&\\
38000 & 1.954(8) & 13917(23)&\\
39000 & 1.983(8) & 15048(24)&\\
\hline
\end{tabular}
\caption{Values of $\gamma$ and $i(v)$ obtained by linear regression analysis. Errors in brackets are standard deviations.}
\label{table1}
\end{table}

\begin{figure}
{\scalebox{0.16}{\includegraphics{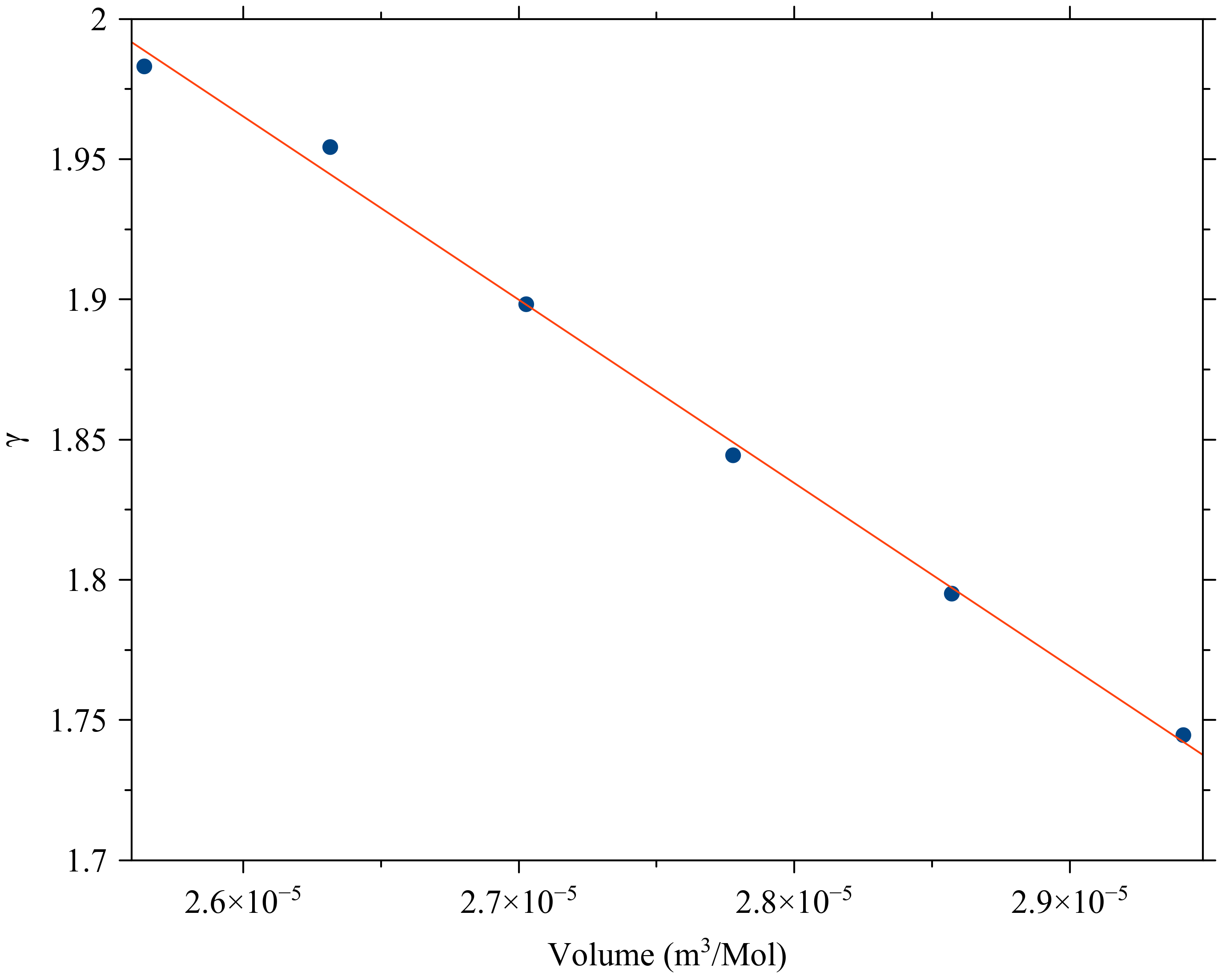}}}
\caption{Values of $\gamma$ obtained from the gradients of $PV$ vs. $E^\prime$ fits to the Ar isochores shown in Figure 1.}
\label{fig2}
\end{figure}

In Table \ref{table1} we see that $\gamma$ itself exhibits a weak, but systematic, volume dependence. This has also been observed for solids. Anderson \cite{anderson1} proposed an empirical volume dependence for solids ($\gamma\propto V^q$ where $q\approx 1$) which has been verified over a wide $P,T$ range. However, the volume dependence in table \ref{table1} for fluid Ar (increase upon volume decrease) is the opposite to that observed for solids. Figure \ref{fig2} shows the trend in $\gamma$ as a function of volume. The trend is roughly linear, and has been fitted using the function $\gamma=\gamma_0+\rho_\gamma V$.

We now proceed to determine the parameters in $E_{st}(V)$ using each approach outlined above. The Lennard-Jones (LJ), constant bulk-modulus ($B$) and pressure-dependent bulk modulus ($B$($P$)) approaches result in the following equations for $i$($V$):

\begin{eqnarray}
\begin{split}
& i_{LJ}(V)=-\gamma E_0+\frac{A}{V^4}(4-\gamma) \\
& i_B(V)=VB_{st}\ln\frac{V}{V_0}(1+\gamma)+VB_{st}-\gamma E_0\\
& i_{B(P)}(V)=-\frac{VB_0}{B^\prime}\Bigl[(1-B^\prime)\left(\frac{V}
{V_0}\right)^{-B^\prime}-1\Bigr]-\\
& \frac{\gamma B_0}{B^\prime}\Bigl[\frac{V^{1-B^\prime}}{V_0^{-B^\prime}}-V\Bigr]-\gamma E_0
\end{split}
\label{j4}
\end{eqnarray}

In Table 1 we see that the obtained values of $i$($V$) increase upon volume decrease. The obtained equation for $i_B$($V$) would require $B_{st}<0$ to reproduce this trend so we can conclude at this stage that the approach of assuming a constant bulk modulus is unsuitable for the study of dense liquids.

In contrast, both $i_{\rm LJ}$($V$) and $i_{B(P)}(V)$ can fit the data from table 1 with physically reasonable values of all fitted parameters. These fits are shown in figure \ref{fig3} and the values of the fitted parameters are:

\begin{figure}
{\scalebox{0.25}{\includegraphics{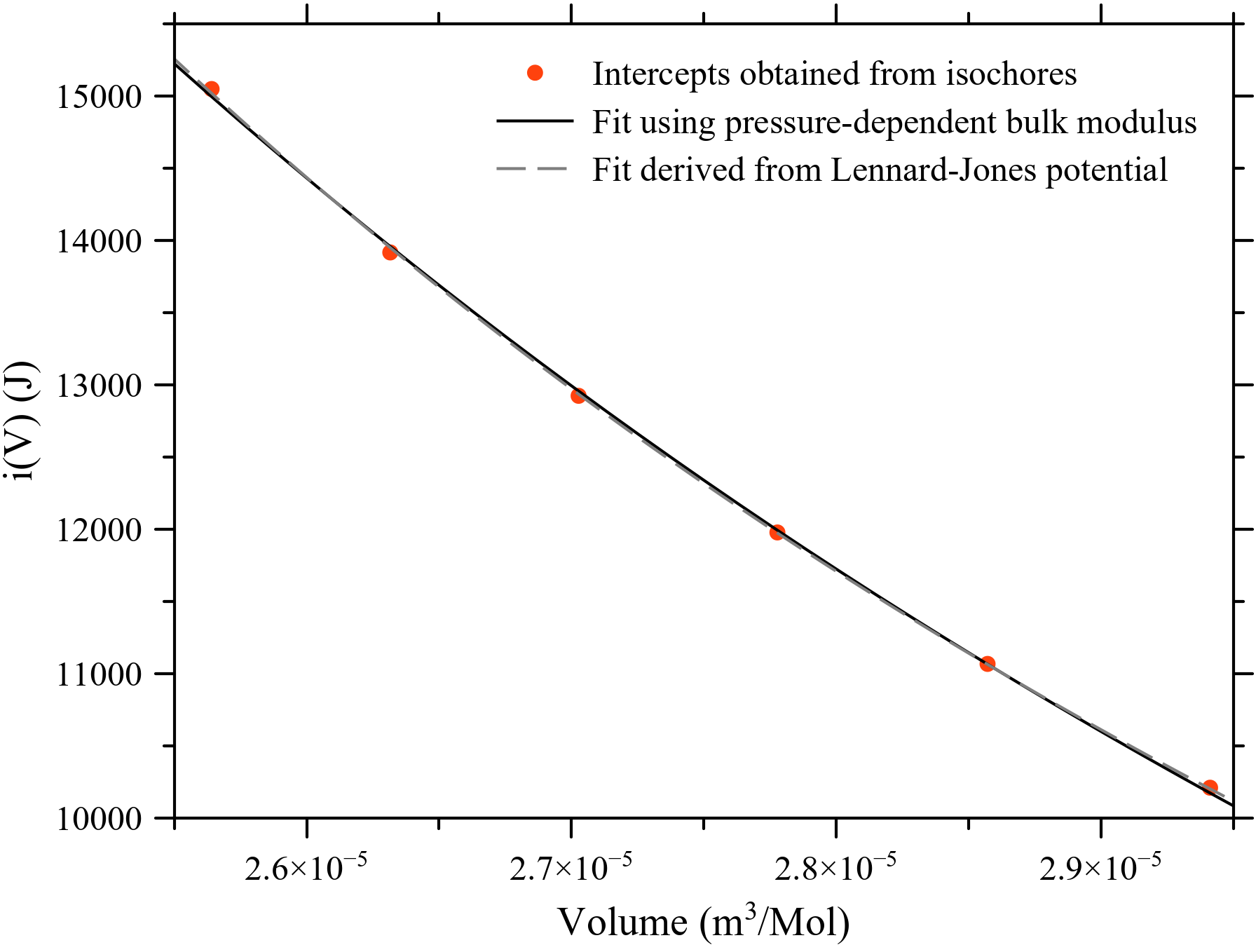}}}
\caption{Plots of the intercepts on the $PV$ axis along various Ar isochores obtained from the experimental data via the fundamental EOS alongside fits using the Lennard-Jones model and pressure-dependent bulk modulus model.}
\label{fig3}
\end{figure}

\begin{eqnarray}
\begin{split}
& A=2.79\times 10^{-15}\pm 2.4\times 10^{-17}~{\rm Jm}^{12}\\
& E_0=-1029\pm 50~{\rm J}\\
& B_0=4.836~{\rm Pa}~{\rm (poorly~constrained)}\\
& V_0=0.0208~{\rm m}^3/\rm{Mol}~(poorly~constrained)\\
& B^\prime=2.10\pm 0.02
\end{split}
\label{j41}
\end{eqnarray}

The fit for $i_{B(P)}(V)$ involves 4 parameters so was poorly constrained (good fits to the data could be obtained using quite different values for the adjustable parameters depending on the choice of initial values). $E_0$ was therefore fixed at the value obtained using our fit to $i_{\rm LJ}$($V$).  This is appropriate since $E_0$ is a property of the experimental data so should not depend on how we choose to fit to it.  Even following this, only $B^\prime$ was well constrained.

\begin{figure}
{\scalebox{0.25}{\includegraphics{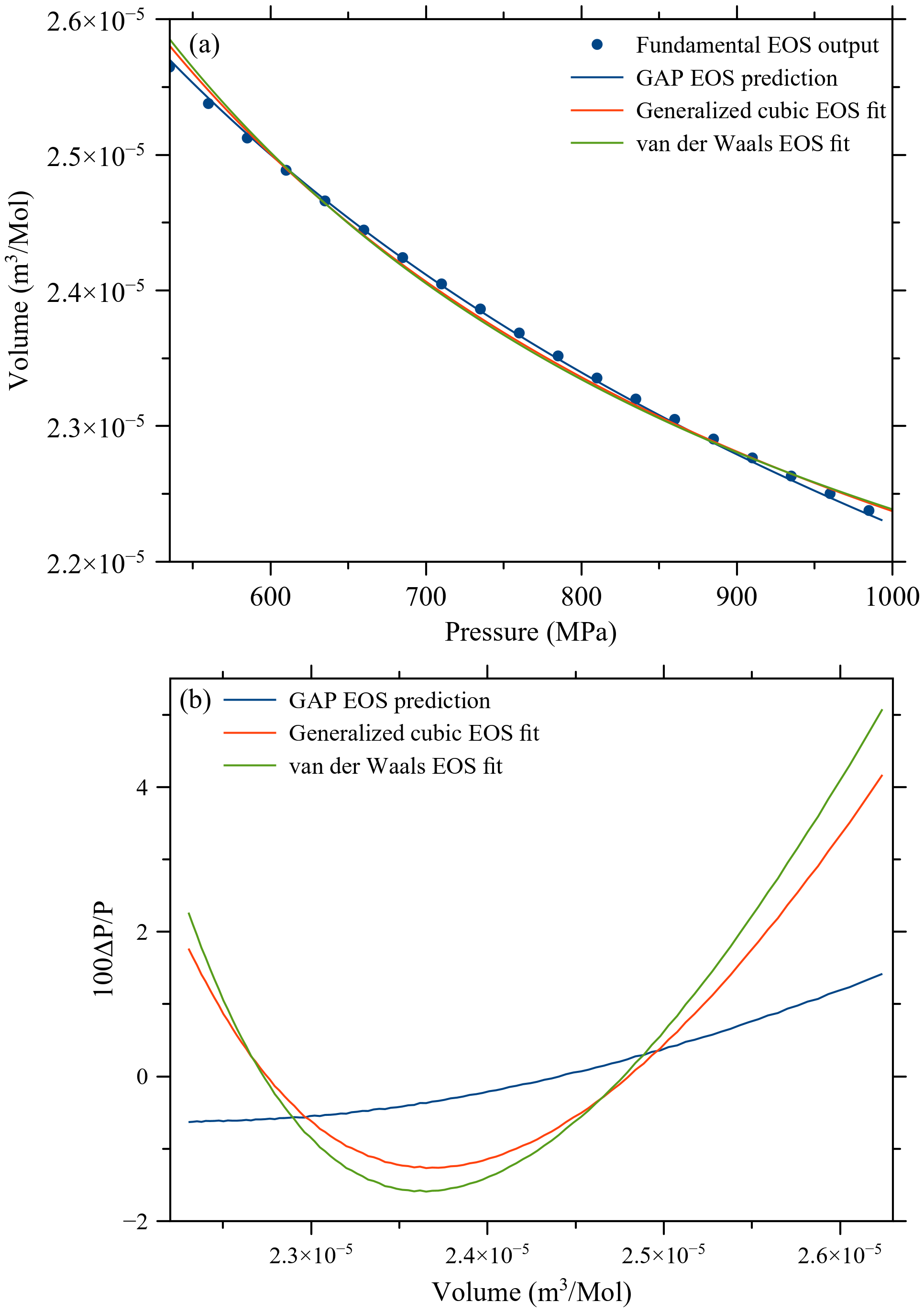}}}
\caption{(a) Fundamental EOS output, extrapolated output from our GAP equation, and fitted van der Waals and generalized cubic PVT EOS at 300 K for fluid Ar from 535 MPa to 1000 MPa. (b) Percentage difference between the pressures calculated using the Fundamental EOS and the GAP equation, van der Waals and generalized cubic PVT EOS for the same volume range as in panel (a).}
\label{fig4}
\end{figure}

At 300 K, our equation has only been fitted to data up to 540 MPa.  We will now examine how well both versions can predict data to which they have not been fitted, using the known $PVT$ EOS of Ar up to 1000 MPa. In Figure \ref{fig4} we show the $PV$ output from the fundamental EOS up to 1000 MPa at 300 K, alongside the $PV$ EOS output from our GAP equation. We present the EOS output using the LJ model since in this case the sole fitting parameter ($A$) is well-constrained. In the pressure range in Figure \ref{fig4} at 300 K our GAP equation fitting parameters have not been fitted to the $PV$ output from the fundamental EOS (compare to Figure \ref{fig2}). The discrepancy between our GAP equation $PV$ output and the fundamental EOS $PV$ output is less than 1\% (this is significantly smaller than the discrepancy between the fundamental EOS output and the original experimental data\cite{robertson}).

The GAP equation performs extremely well compared to cubic EOS.  For comparison, we have fitted the van der Waals PVT EOS and generalized cubic PVT EOS to the fundamental EOS output for Ar to 1000 MPa at 300 K.  As we can see from figure \ref{fig4}, the prediction with the GAP equation is a far better fit to the data than the actual fit using the van der Waals EOS (equation \eqref{vderw}) or generalized cubic EOS (equation \eqref{gencub}).

In the theory section, we outlined an alternate approach: in a wide part of the phase diagram where the hopping frequency $\omega_{\rm F}$ is much smaller than the Debye frequency, $\omega_{\rm D}$, $E_{th}=3NT$ (see Eq. \eqref{harmo}).  Using this approach, our GAP equation can be tested without referring to the internal energy output from the fundamental EOS and simplifies to equation \eqref{eosl1}.

\begin{figure}
{\scalebox{0.25}{\includegraphics{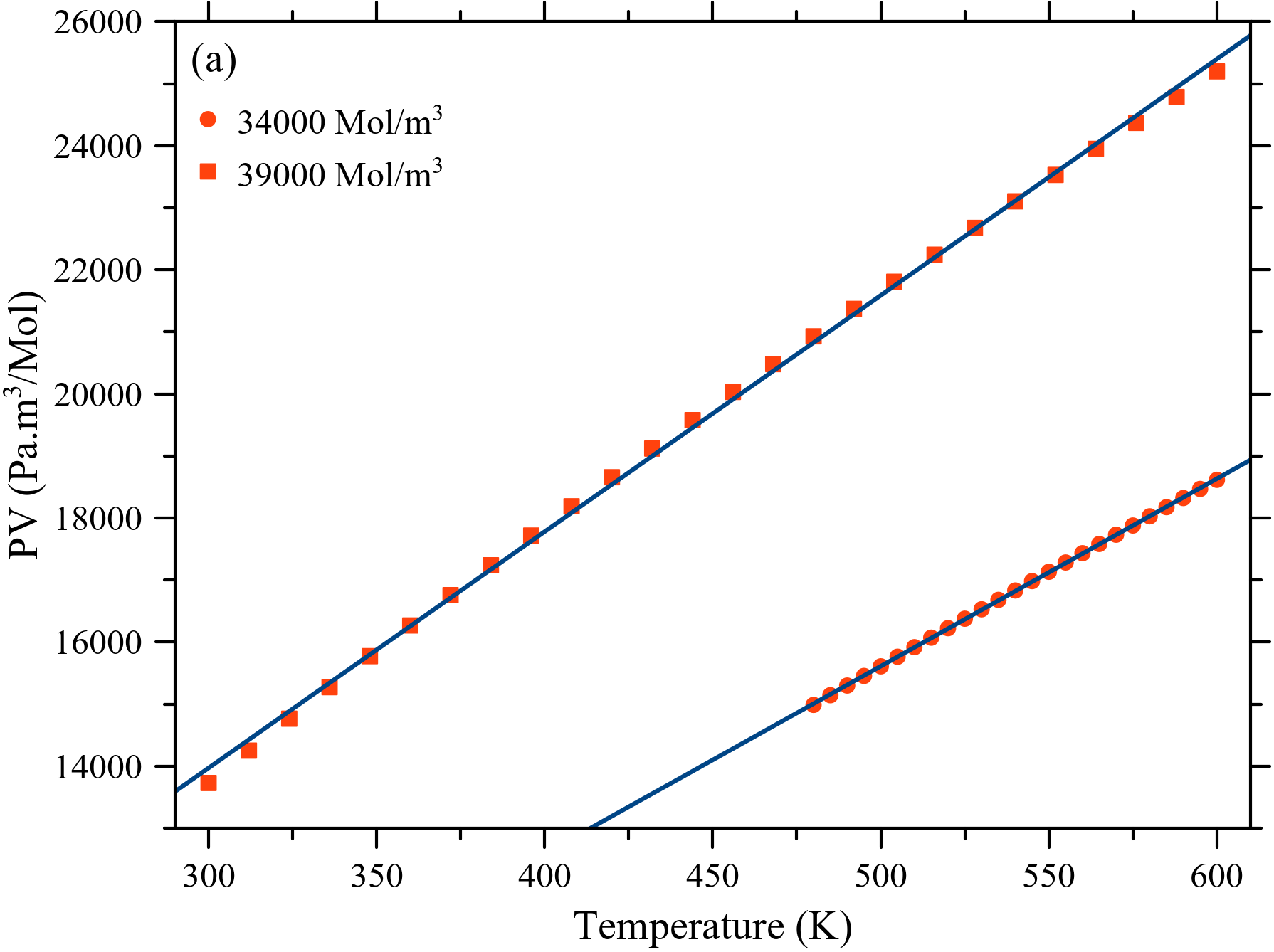}}}
{\scalebox{0.25}{\includegraphics{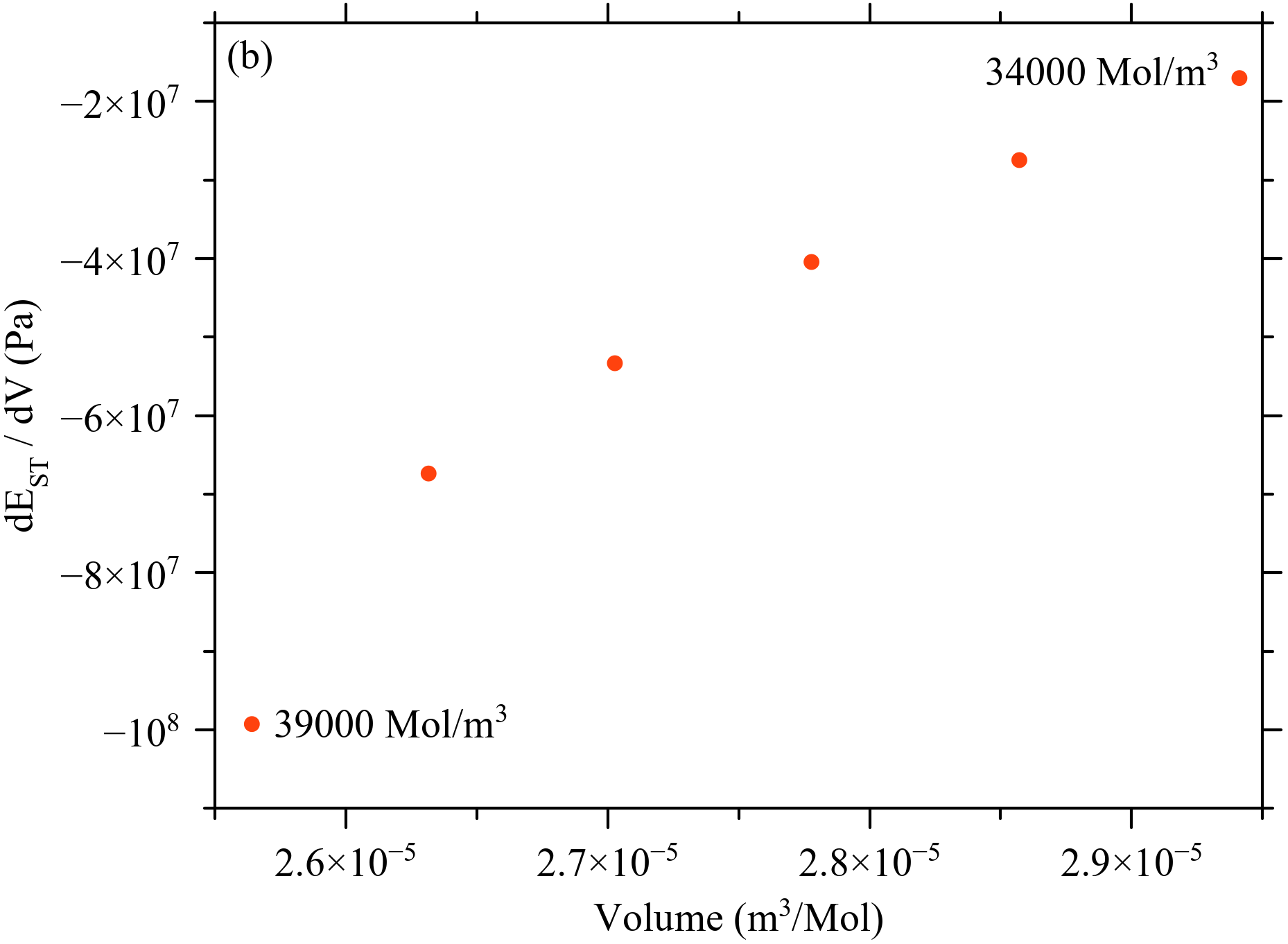}}}
\caption{(a) Plots of $PV$ versus $T$ for the lowest and highest Ar isochores studied, with linear fits. (b) Values of $\frac{dE_{st}}{dV}$ obtained from the intercepts of these linear plots for all Ar isochores.}
\label{fig5}
\end{figure}

In this case a plot of $PV$ versus $T$ along an isochore should be linear, with a gradient of $3\gamma N$. Figure \ref{fig5} shows the plots with linear fits for the lowest and highest density isochores studied, and the trend in the gradient upon increasing density. The trend would correspond to a variation in the value of $\gamma$ of about 20\% between the lowest and highest density isochore (a larger variation than that found for the other model earlier in this section but not unreasonable).

The fact that the linear relation is observed along isochores indicates that this implementation of our GAP equation is physically sound, and the intercept of this graph provides values of $\frac{dE_{st}}{dV}$ at each of the six densities studied. The trend is physically realistic. However, a good fit with physically realistic values of the fitting parameters cannot be obtained with either the Lennard-Jones model or the pressure-dependent bulk modulus model.

The agreement of the GAP equation and the experimental isochoric data in liquids is important because it involves a model-free way to test the GAP EOS. It therefore carries the same significance as testing the M-G EOS in solids, one of the most important and common EOS used in the solid state theory \cite{anderson}.

\subsection{Testing the new EOS against experimental data for N$_2$}

\begin{figure*}
{\scalebox{0.35}{\includegraphics{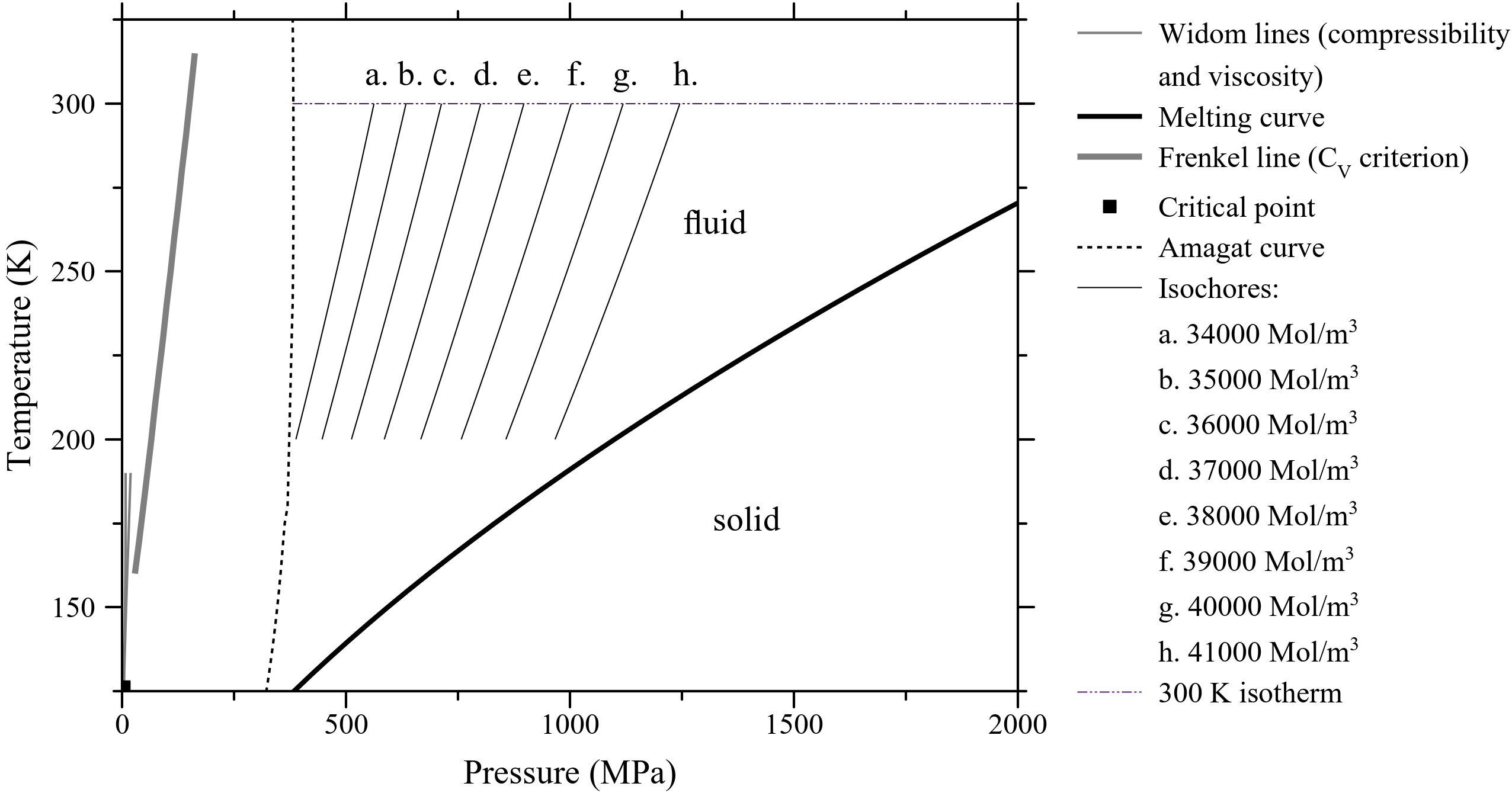}}}
\caption{Pressure-temperature phase diagram of N$_2$ produced using the methodology outlined in Ref. \cite{proctor2}. $P,T$ paths along which our GAP equation has been fitted and tested are marked.}
\label{fig6}
\end{figure*}

Testing with molecular fluids poses the additional complication that intra-molecular modes contribute to the internal energy.  In the case of N$_2$, in the temperature range 200-300 K there is adequate thermal energy available to excite the rotational mode but not the vibrational mode so this can be accounted for by subtracting $RT$ from the internal energy $E^\prime$.  The fundamental EOS internal energy output is backed by speed of sound and $PVT$ data up to the freezing point (ca. 1100 MPa) at 200 K and to 2000 MPa at 300 K. A wide range of testing can therefore take place along isochores and isotherms. Figure \ref{fig6} shows the phase diagram of N$_2$. The isochores used to fit our GAP equation to the experimental data, and the isotherm along which it has been tested, are marked.

\begin{figure}
{\scalebox{0.17}{\includegraphics{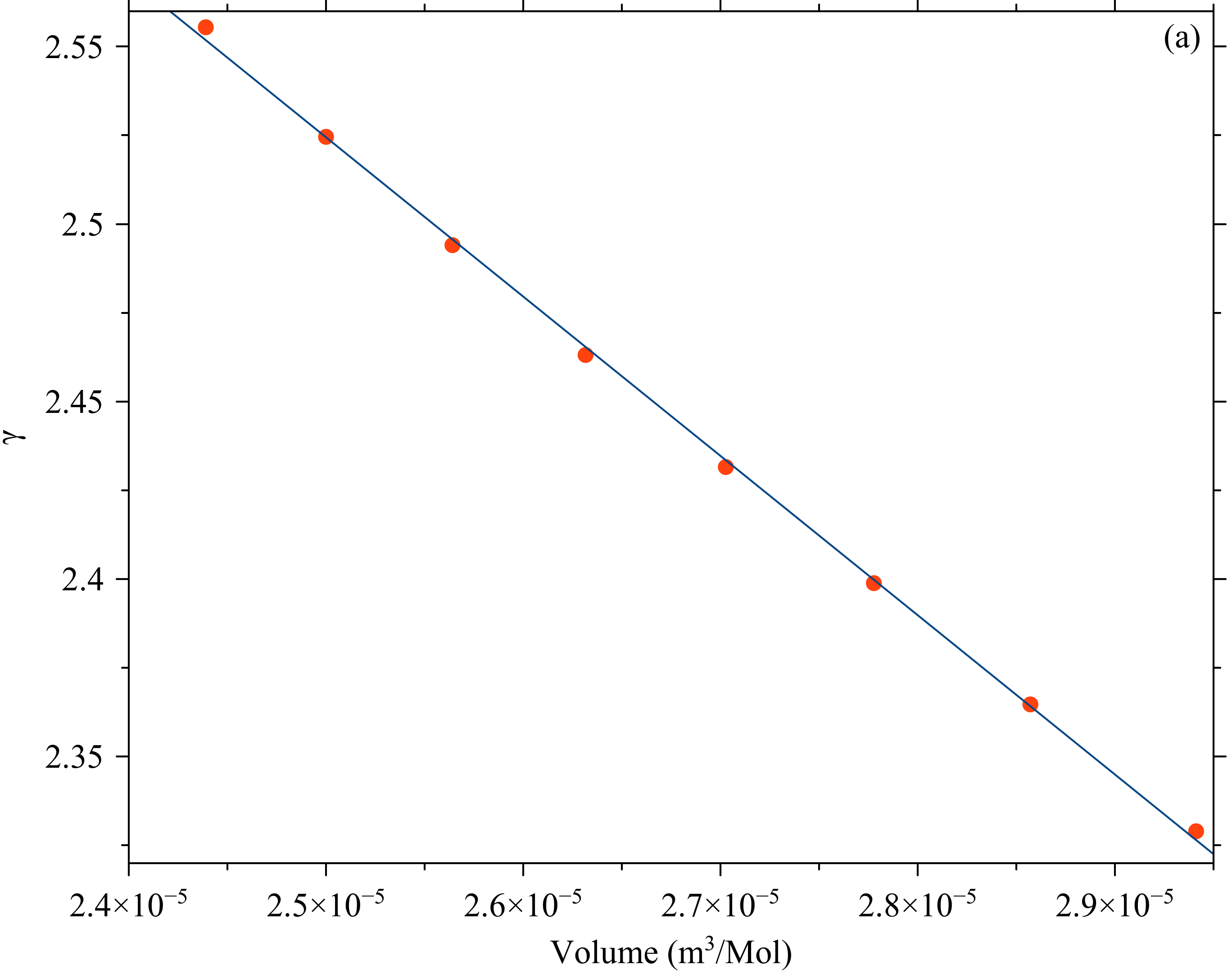}}}
{\scalebox{0.17}{\includegraphics{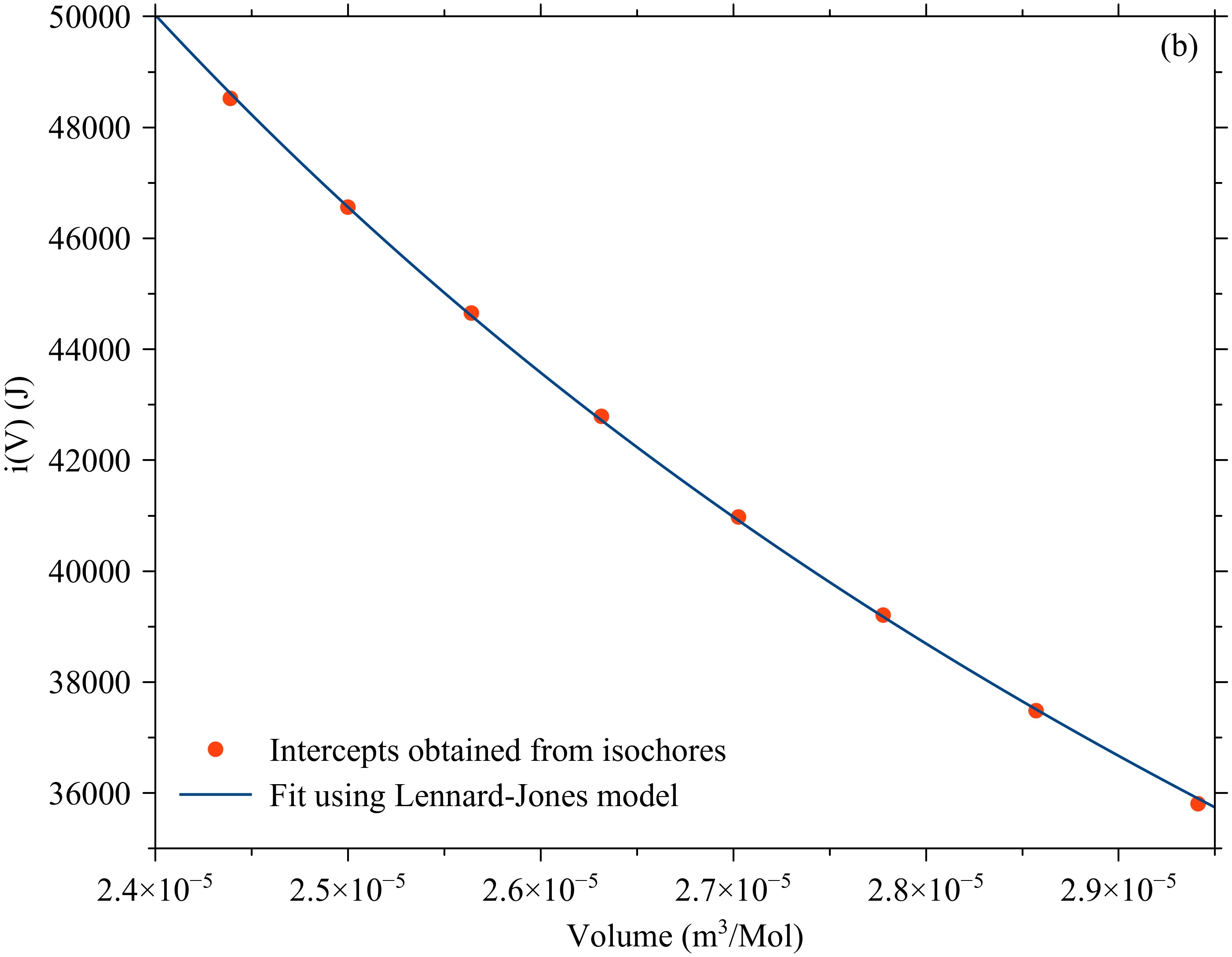}}}
\caption{Fit parameters from application of GAP equation to N$_2$ in $PVE$ form. (a) Plot of $\gamma$ versus volume for N$_2$ isochores with linear fit. (b) Plot of intercept $i$ versus volume for N$_2$ isochores with fit using LJ model.}
\label{fig7}
\end{figure}

\begin{figure}
{\scalebox{0.26}{\includegraphics{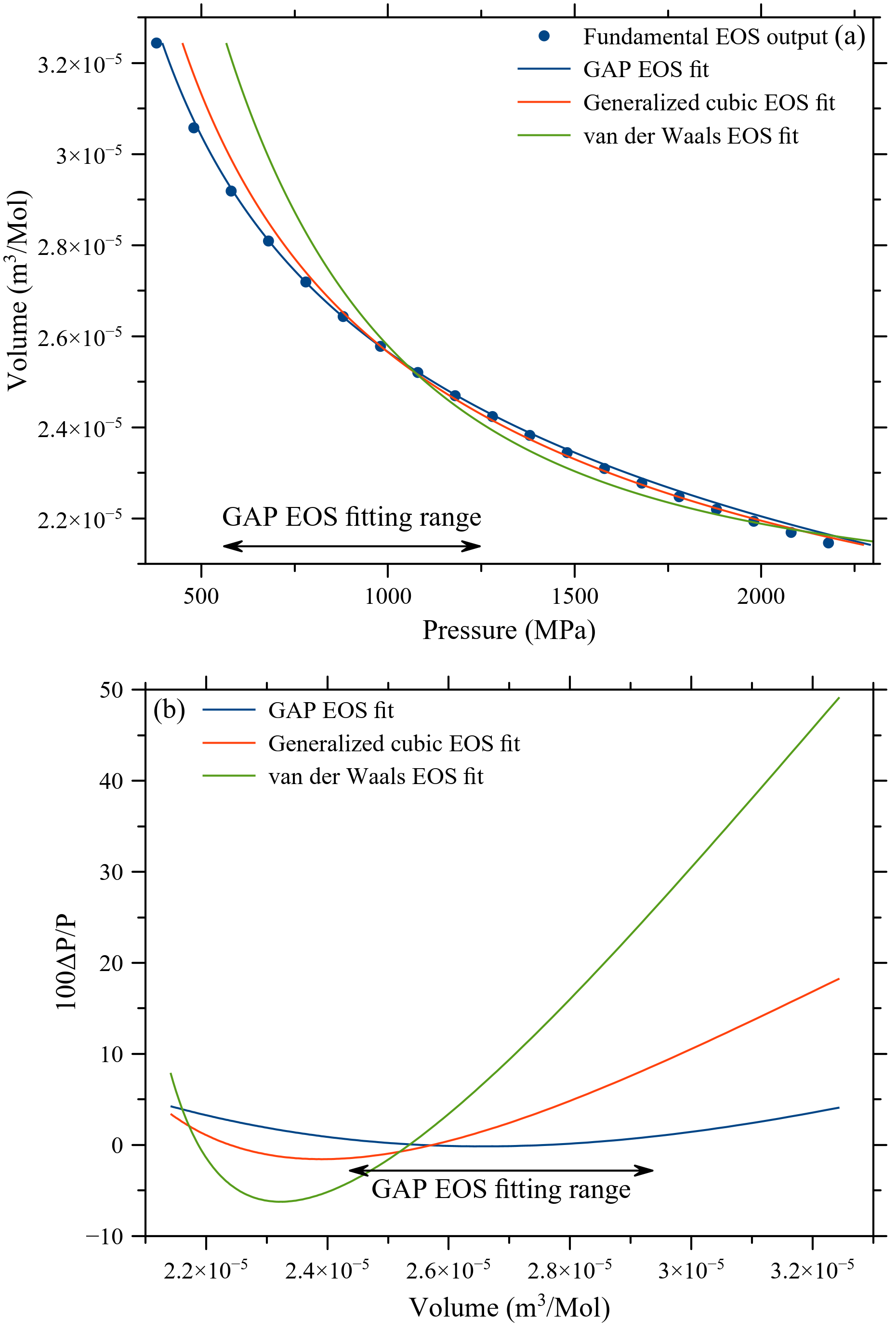}}}
\caption{(a) Fundamental EOS output and output from our GAP equation, along with the generalized cubic and van der Waals PVT EOS for fluid N$_2$ from 400 MPa to 2200 MPa at 300 K. The pressure range in which our GAP equation was fitted to the fundamental EOS output is marked.  (b) Percentage difference between the pressures calculated using the Fundamental EOS and the GAP equation, van der Waals and generalized cubic PVT EOS for the same volume range as in panel (a).}
\label{fig8}
\end{figure}

\begin{figure}
{\scalebox{0.15}{\includegraphics{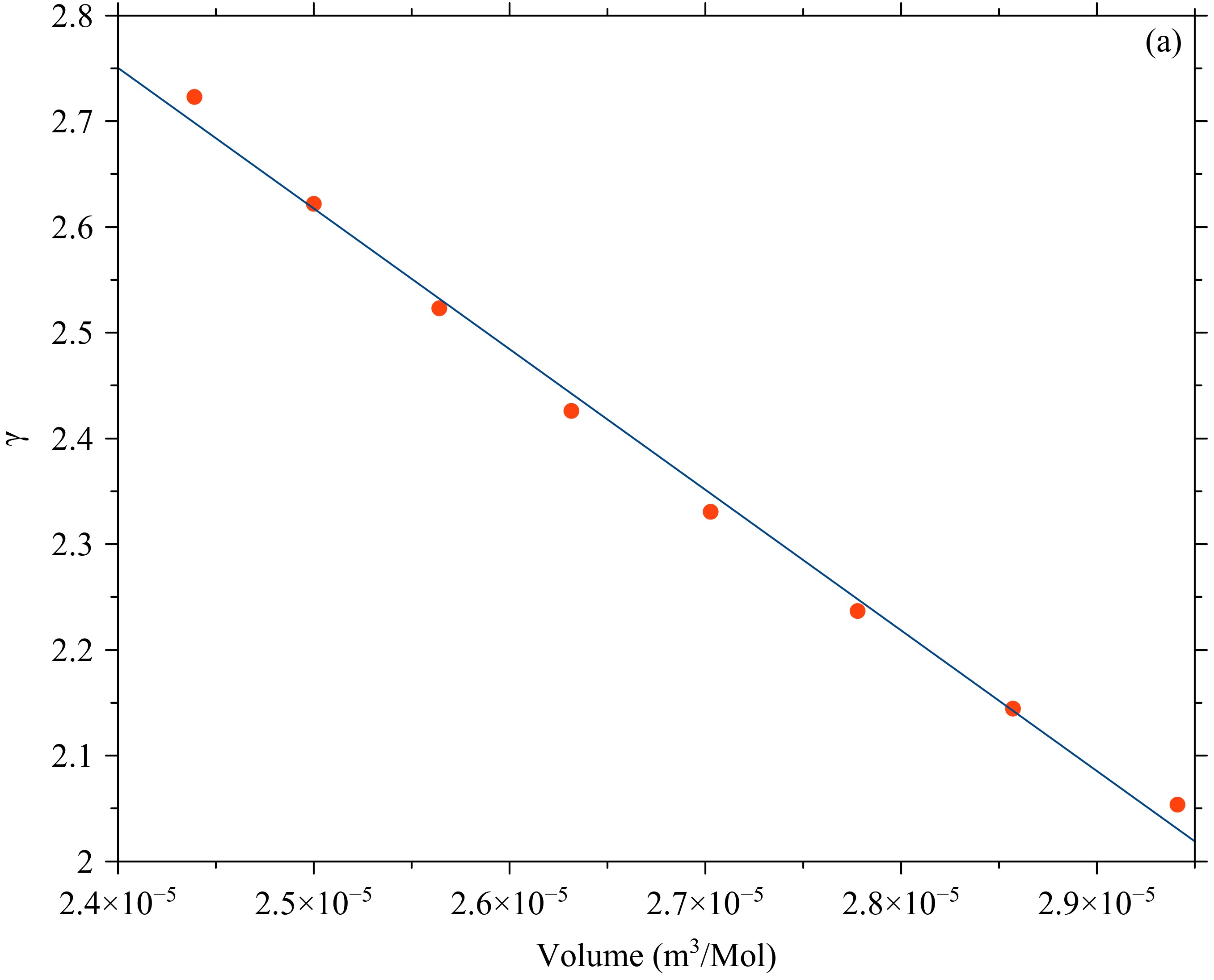}}}
{\scalebox{0.25}{\includegraphics{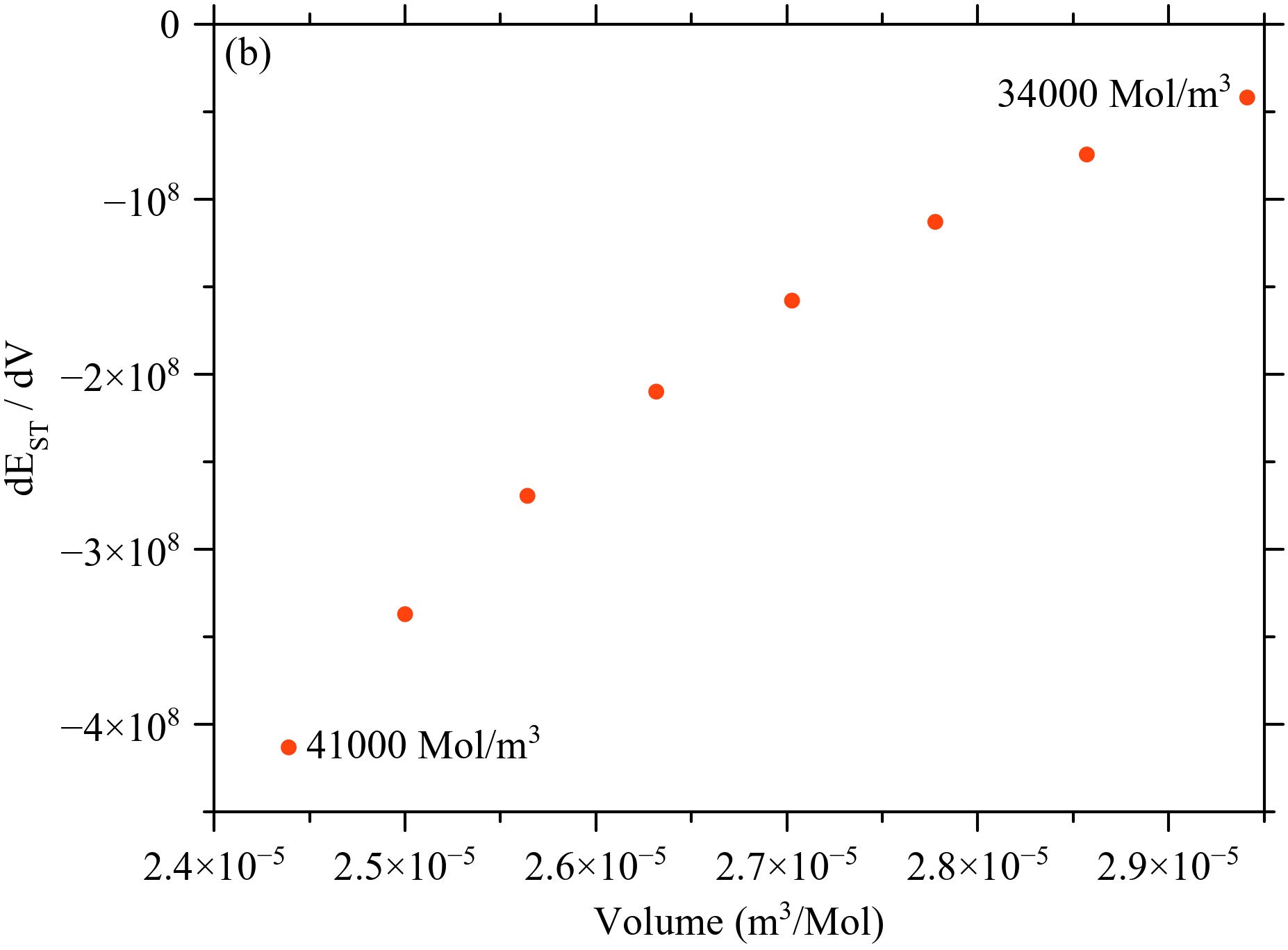}}}
\caption{Fit parameters from application of GAP Equation to N$_2$ in $PVT$ form. (a) Trend in $\gamma$ as a function of volume for N$_2$ along the isochores studied. (b) Trend in $\frac{dE_{st}}{dV}$ as a function of volume for the same isochores.}
\label{fig9}
\end{figure}

Similarly to Ar, we obtained a linear $PV$ vs $E^\prime$ relation along all the isochores studied and obtained the trend in $\gamma$ (fitted with $\gamma=\gamma_0+\rho_\gamma V$ in figure \ref{fig7}a). The intercept $i$ was plotted analogously to Ar (figure \ref{fig7}b).  In this case the fits using the LJ model and pressure-dependent bulk modulus model produced very similar trends so we have shown only the result with the LJ model.

Similarly to Ar, our fitting along isochores allows us to then test our GAP equation by predicting the pressure at smaller volume along the isotherm shown in figure \ref{fig6} (300 K from the Amagat curve up to the limit of the original experimental data at 2000 MPa).  Figure \ref{fig8} shows the results of this testing: prediction of the volume to within 1\% at the highest pressure studied, similar to the error between the fundamental EOS\cite{spanN2EOS} and the original experimental data \cite{mills}.  This is in contrast to the van der Waals and generalized cubic PVT EOS, which perform very poorly even when fitted to the entire range of PVT data.

Our final investigation on fitting our GAP equation to the fluid N$_2$ data is to investigate how we can fit to the experimental data assuming $E_{th}=3NT$, leading to equation \eqref{eosl1}. We found that the graphs of $PV$ vs $T$ along all isochores were linear to a good approximation. The value of $\gamma$ can be obtained from the gradient and the value of $\frac{dE_{st}}{dV}$ from the intercept. Figure \ref{fig9} shows the observed parameters and their trends as a function of volume. The parameters and their trends are physically reasonable. The trend in $\gamma$ can be fitted linearly, whilst neither the LJ or the pressure-dependent bulk modulus model can fit the trend in $\frac{dE_{st}}{dV}$ whilst retaining physically reasonable values of all fitting parameters. These outcomes are similar to Ar.

To summarise this section, we used several different methods to test the predictions of our GAP equation against experimental data for Ar and N$_2$. We saw that each method returns the result that the GAP equation is predictive and consistent with the experimental data.

Due to its’ derivation from the phonon theory of liquid thermodynamics, the GAP EOS is applicable on the rigid liquid (higher pressure) side of the Frenkel line.  In the testing conducted so far, we have utilized a function for $E_{ST}(V)$ which accounts for repulsion (but not attraction) between fluid particles.  In this case, the applicability of the equation is also constrained by the Amagat curve (the equation is applicable on the high pressure side of the Amagat curve).

\section{Discussion and conclusions}
\label{discussion}

The agreement between our GAP equation and the available experimental data indicates that it can in future find widespread applications in condensed matter physics, chemical engineering and planetary science.  The agreement, combined with the failure of the constant bulk modulus model, and differences between our findings for liquids and existing knowledge on EOS of solids, give confidence that the success of our GAP equation is due to it being grounded in the laws of physics, and is not merely a serendipity. We note in particular the following points.

First, the difference between the constant bulk modulus model and (on the other hand) the Lennard-Jones and pressure-dependent bulk modulus models is that the former does not make any attempt to account for the repulsion between atoms varying with density. If integrated, the constant bulk modulus model results in an exponential decay in volume as a function of pressure. So $V\rightarrow 0$ is permitted and atoms are allowed to overlap. The fact that this model is unable to fit the data whilst the Lennard-Jones and pressure-dependent bulk modulus models both fit the data extremely well, is what we would expect for fluids at densities approaching close-packed structures.

Second, the values obtained for $B^\prime$ (ca. 2 in both cases) are exactly as would be expected. For solids, $B^\prime\approx 4$ is typically obtained. Since liquids are not quite as closely-packed as solids, we would expect a model incorporating a slightly weaker pressure dependence of the bulk modulus to work best.

Third, the implementation of our GAP equation with the approximation $E_{th}=3NT$ and ensuing linearity observed along the $PV$ vs $T$ plots for isochores supports the approximation we made in simplifying our GAP equation in Eq. \eqref{eos1} or Eq. \eqref{eos2} to result in Eq. \eqref{eosl1} and \eqref{eosl2}. This EOS can potentially be used in cases where no internal energy data are available, or find future application for fluids at high density and/or high viscosity. We note that, with $E_{th}=3NT$, the trend in $\frac{dE_{st}}{dV}$ was physically realistic but neither the LJ or pressure-dependent bulk modulus model could fit to it. So our GAP equation - derived from first principles - is working, but the empirical models for $E_{ST}(V)$ are not. Clearly, an important avenue for further development of the GAP equation is to combine it models for $E_{ST}(V)$ that are based more closely on first principles than those used in the present work.

Fourth, we recall that $\gamma(V)$ shows the opposite trend in liquids and solids. The liquid thermal energy $E_{th}$ (in contrast to solids) has a contribution which decreases with volume according to Eq. \eqref{harmo} because the hopping frequency $\omega_{\rm F}$ set by viscosity increases with volume. This contributes to the decrease of the thermal pressure in Eq. \eqref{eosl2} $P_{th}=\frac{1}{V}\gamma E_{th}$, with the consequence for the fitted $\gamma$. The fact that $\gamma(V)$ displays the same trend in all testing done so far suggests that it should be incorporated into the GAP equation. For future use, the GAP equation can therefore be written with $\rho_\gamma<0$ as:

\begin{equation}
PV=-V\frac{dE_{st}}{dV}+3(\gamma_0+\rho_\gamma V)NT
\label{eosl11}
\end{equation}

\begin{equation}
PV=-V\frac{dE_{st}}{dV}+(\gamma_0+\rho_\gamma V)E_{th}
\label{eosl22}
\end{equation}

Fifth, recall that the GAP equations, once approximated using $\omega_{\rm F}\ll\omega_{\rm D}$, become close to the M-G EOS for solids. This has consequences for a more general outlook at liquids as the third basic state of matter (in addition to solids and gases). Similarly to gases and differently from solids, liquids flow (this flow ability has given rise to hydrodynamic approaches to liquids based on Navier-Stokes and related equations, the approach that has been limited in capturing some essential liquid properties because it missed the solidlike vibrational properties discussed in references \cite{ropp,mybook}). However, this similarity between liquids and gases in terms of the ability to flow is pretty much where the essential similarities end. Many key properties such as density, compressibility, elastic moduli, thermal conductivity as well as internal energy and heat capacity close to melting (see Section \ref{liquidexc}) are very similar in solids and liquids but are very different in liquids and gases (as discussed earlier, thermodynamic properties such as energy and heat capacity of liquids and solids are very close in a wide range of the liquid phase diagram where $\omega_{\rm F}\ll\omega_{\rm D}$). For this reason, the EOS, the consequence of thermodynamic properties, is expectedly similar in solids and liquids in that part of the phase diagram and is very different in liquids and gases as mentioned earlier.

Fundamentally, this similarity stems from the fact that liquids and solids are condensed states of matter, whereas gases are not. This condensed state means that the scale of energy and length in liquids are set by their characteristic values related to the Rydberg energy and Bohr radius as they are in solids \cite{myreview}. The similarity between solids and liquids in terms of elastic, thermal, transport and other essential properties such as sound propagation then follows (see Ref. \cite{myreview} for review). On the other hand, these energy and length scales are inapplicable to gases where atoms are not affected by cohesion and where the Bohr length scale and cohesive Rydberg energy scale do not operate. As a result, the ensuing properties listed above are very different in gases.

In addition to better understanding the high-temperature liquid state, the above closeness between the liquid and solid EOS is useful for understanding the low-temperature viscous regime and the glass transformation range. This is the area that has been of interest for many decades and is still developing \cite{dyre,ropp,mybook}. As discussed in Section \ref{twoforms}, the condition $\omega_{\rm F}\ll\omega_{\rm D}$ and ensuing Eq. \eqref{eosl} applies to the very viscous liquids particularly well. Hence, our EOS and its predictions can be reliably applied to liquids in the glass transformation range. Our EOS can also be extended to other systems such as solidlike phases in soft matter. For example, this includes packing problems in soft and granular matter systems where model liquid EOS are useful to understand packing fractions \cite{zacpac}.

In our previous work on the phonon theory of liquid thermodynamics \cite{ropp,proctor1,proctor2} we began with a careful look at liquid viscosity and related the liquid viscosity to the key microscopic property of the liquid (the intermolecular hopping frequency $\omega_{\rm F}$ or liquid relaxation time $\tau$). This allowed us to construct an $E_{TH}\eta$ EOS for liquids with only two fitting parameters: the Debye wavenumber and the infinite-frequency shear modulus. These took physically realistic values. Even the heat capacity trends could be correctly reproduced by allowing only tiny adjustments to liquid relaxation time, within the error margin of the viscosity data from which it was calculated.
In the present work, we proceed from $E_{TH}$ to directly link the most fundamental macroscopic property of fluids, the $PVT$ EOS, directly to the key microscopic property of the liquid relaxation time.  The direct application of the M-G EOS to liquids was proposed some time ago \cite{amoros}, but it's proposal was purely a consequence of phenomenological observations based on very limited data available at the time.  Here, using the phonon theory of liquid thermodynamics we are able to establish that the application of an EOS (the GAP EOS) related to the M-G EOS to liquids comes directly from first principles.  We are able to demonstrate quantitative agreement with experimental data over a wide $PT$ range in which data did not exist in the 1980s.

The link from viscosity (or, equivalently, hopping frequency $\omega_{\rm F}$ to internal energy and heat capacity in Refs. \cite{ropp,proctor1,mybook} and to the PVT EOS (present work) via the liquid relaxation time is an advance in itself as it allows these parameters to be fitted, and predicted, via a single model when previously they were treated using separate models.  The fundamental equation of state provides static properties (e.g. the PVT EOS) and dynamic properties (e.g. heat capacity) from the same model but cannot provide transport properties (viscosity, thermal conductivity) due to the fact that it is explicit in the Helmholtz free energy $F(V,T)$, and there is no thermodynamic relationship that allows transport properties to be obtained from this.  Transport properties have, until now, been provided by a completely separate mainly empirical model (e.g. ref. \cite{lemmonvisc}).

One limitation of the GAP equation is it's applicability to more complex fluids over a wide $PT$ range, which may be limited due to the need to subtract the contribution from intra-molecular modes to the internal energy.  We performed this subtraction earlier for N$_2$ but it would become more complex for polyatomic molecules.  Applying the GAP equation to polyatomics may therefore require models describing the contribution to the internal energy from intra-molecular modes, or a restriction to working along isotherms.

We note that there is still a critical lack of experimental data on hot dense fluids in particular, leading to the fitting parameters in our GAP equation being poorly constrained in many cases. To advance the field further, it is necessary to fit using data analogous to Figures 2 and 3 over a wider range of experimental conditions. Primarily this is because there is a gap (especially at high temperature) between the maximum pressures obtainable in piston-cylinder devices offering direct volume measurement and the melting curves for noble gases and simple systems such as Ar and N$_2$. There is a critical lack of $PVT$ EOS data in the literature on hot fluids measured in the DAC (although some promising progress has been made very recently \cite{JPDACEOS}). This gap needs to rectified, and we hope that our discussion will serve as an additional stimulus to pursue these experiments.

\section{Acknowledgement}

We would like to acknowledge useful feedback from Vladimir Diky (NIST) and Gijsbertus De With (Eindhoven University of Technology).

%

\bibliographystyle{apsrev4-1}

\end{document}